\documentclass{aa}  
\usepackage{txfonts}
\usepackage{hyperref}

\usepackage{graphicx}	
\usepackage{amsmath}	
\usepackage{amssymb}	

\newcommand*{\unit}[1]{~\mathrm{#1}}
\newcommand*{\lunit}{\ensuremath{\mathrm{erg\,s^{-1}}}}
\newcommand*{\funit}{\ensuremath{\mathrm{erg\, cm^{-2}s^{-1}}}}
\newcommand*{\xmm}{\textit{XMM-Newton}}
\newcommand*{\chandra}{\textit{Chandra}}
\newcommand{\angstrom}{\mbox{\normalfont\AA}}
\newcommand{\daox}{\mbox{$\Delta\alpha_{ox}$}}

\begin{document} 
\title{The X-ray properties of $z>6$ quasars: no evident evolution of accretion physics in the first Gyr of the Universe}
\titlerunning{The X-ray properties of $z>6$ QSOs}

\authorrunning{F. Vito et al.}
   \author{F. Vito\thanks{fabio.vito@uc.cl}\inst{1,2,3,4} \and
          W.N. Brandt\inst{3,4,5} \and
          F.E. Bauer\inst{1,6,7} \and
          F. Calura\inst{8} \and
          R. Gilli\inst{8} \and
          B. Luo\inst{9,10} \and
          O. Shemmer\inst{11} \and
          C. Vignali\inst{8,12} \and
          G. Zamorani\inst{8} \and
          M. Brusa\inst{8,12} \and
          F. Civano\inst{13} \and
          A. Comastri\inst{8} \and
          R. Nanni\inst{8,12}
          }
      
    \institute{Instituto de Astrofisica and Centro de Astroingenieria, Facultad de Fisica, Pontificia Universidad Catolica de Chile,\\ Casilla 306, Santiago 22, Chile 
    	\and
      Chinese Academy of Sciences South America Center for Astronomy, National Astronomical Observatories, CAS, Beijing 100012, China
      \and
     Department of Astronomy \& Astrophysics, 525 Davey Lab, The Pennsylvania State University, University Park, PA 16802, USA
           \and
      Institute for Gravitation and the Cosmos, The Pennsylvania State University, University Park, PA 16802, USA
            \and
      Department of Physics, The Pennsylvania State University, University Park, PA 16802, USA
            \and
      Millennium Institute of Astrophysics (MAS), Nuncio Monse\~nor S\'otero Sanz 100, Providencia, Santiago, Chile
            \and
      Space Science Institute, 4750 Walnut Street, Suite 205, Boulder, Colorado, 80301, USA
            \and
    INAF -- Osservatorio di Astrofisica e Scienza dello Spazio di Bologna, Via Gobetti 93/3, I-40129 Bologna, Italy
            \and
   School of Astronomy and Space Science, Nanjing University, Nanjing 210093, China
            \and
     Key Laboratory of Modern Astronomy and Astrophysics (Nanjing University), Ministry of Education, Nanjing, Jiangsu 210093, China
            \and
    Department of Physics, University of North Texas, Denton, TX 76203
            \and
    Dipartimento di Fisica e Astronomia, Universit\`a degli Studi di Bologna, via Gobetti 93/2, 40129 Bologna, Italy 
            \and
   15 Center for Astrophysics | Harvard \& Smithsonian, 60 Garden St, Cambridge, MA 02138, USA
}

   \date{}
  \abstract
   {X-ray emission from quasars (QSOs) has been used to assess supermassive black hole (SMBH) accretion properties up to $z\approx6$. However, at $z>6$ only $\approx15$ QSOs are covered by sensitive X-ray observations, preventing a statistically significant investigation of the X-ray properties of the QSO population in the first Gyr of the Universe.}
   {We present new \chandra\, observations of a sample of 10 $z>6$ QSOs, selected to have virial black-hole mass estimates from Mg II line spectroscopy (log$\frac{M_{\mathrm{BH}}}{M_\odot}=8.5-9.6$). Adding archival X-ray data for an additional 15 $z>6$ QSOs, we investigate the X-ray properties of the QSO population in the first Gyr of the Universe, focusing in particular on the $L_{UV}-L_{X}$ relation, which is traced by the $\alpha_{ox}$ parameter, and the shape of their X-ray spectra.}
   {We performed photometric analyses to derive estimates of the X-ray luminosities of our $z>6$ QSOs, and thus their $\alpha_{ox}$ values and bolometric corrections ($K_{bol}=L_{bol}/L_{X}$). We compared the resulting $\alpha_{ox}$ and $K_{bol}$ distributions with the results found for QSO samples at lower redshift, and ran several statistical tests to check for a possible evolution of the $L_{UV}-L_{X}$ relation. Finally, we performed a basic X-ray spectral analysis of the brightest $z>6$ QSOs to derive their individual photon indices, and joint spectral analysis of the whole sample to estimate the average photon index.}
   {We detected seven of the new \chandra\, targets in at least one standard energy band, while two more are detected discarding energies $E>5$ keV, where background dominates. We confirm a lack of significant evolution of $\alpha_{ox}$ with redshift, extending the results from previous works up to $z>6$ with a statistically significant QSO sample, and the trend of an increasing bolometric correction  with increasing luminosity found for QSOs at lower redshifts. The average power-law photon index of our sample ($\langle\Gamma\rangle=2.20_{-0.34}^{+0.39}$ and $\langle\Gamma\rangle=2.13_{-0.13}^{+0.13}$ for sources with $<30$ and $>30$ net counts, respectively) is slightly steeper than, but still consistent with, typical QSOs at $z=1-6$.}
   {All these results point toward a lack of substantial evolution of the inner accretion-disk/hot-corona structure in QSOs from low redshift to $z>6$. Our data hint at generally high Eddington ratios at $z>6$.}


   \maketitle
%

\section{Introduction }\label{intro}

X-ray emission from accreting supermassive black holes (SMBHs), shining as quasars (QSOs), is thought to originate from inverse Compton scattering in the so-called ``hot corona" of the UV/optical photons produced by the accretion disk via thermal emission \citep[e.g.][]{Galeev79,Haardt91,Beloborodov17}.
The relative importance of the  hot corona and the accretion disk to the total radiative output is usually parametrized with \hbox{$\alpha_{ox}=0.38\times \mathrm{log}(L_{2\,\mathrm{keV}}/L_{2500\angstrom})$}, which \mbox{represents} the slope of a nominal power-law connecting the rest-frame UV and \mbox{X-ray} emission (e.g. \citealt{Brandt15} and references therein). $\alpha_{ox}$ is known to anti-correlate with the QSO UV luminosity \citep[e.g.][see \citealt{Lusso17} for a physical interpretation]{Steffen06,Just07,Lusso16}, i.e. the fractional disk contribution to the total emitted power increases for more luminous QSOs. Most previous works \citep[e.g.][]{Vignali03, Steffen06,Just07,Jin12, Marchese12,Lusso16, Nanni17} found no evidence for evolution with redshift of $\alpha_{ox}$. Recently, \mbox{\cite{Risaliti19}} exploited this apparent lack of evolution to propose QSOs as standard candles to infer cosmological parameters (see also Salvestrini et al., submitted).

Currently, $\approx190$ QSOs have been discovered at \mbox{$z\ge6$}, corresponding to about the first Gyr of the Universe (\citealt{Banados16} and references therein; \citealt{Mazzucchelli17b,Reed17,Reed19,Tang17,Wang17,Wang18a,Wang18b,Chehade18,Matsuoka18a,Matsuoka18b,Matsuoka19, YangJ18,Fan19, Pons19}), with ULASJ1342+0928 holding the redshift record of $z=7.54$ \citep{Banados18a}.
These rare QSOs were selected in wide-field optical/near-IR surveys such as SDSS, CFHQS, UKIDSS, Pan-STARRS1, ATLAS, and VIKING, and represent the extreme tail of the underlying SMBH population at early epochs. For instance, most of the known $z>6$ QSOs are extremely luminous (log$L_\mathrm{bol}/L_\odot\approx12-14$) and massive (up to $\approx10^{10}\,M\MakeLowercase{_\odot}$; \citealt{Wu15}).
The very existence of such massive black holes in the early universe challenges our theoretical knowledge of SMBH formation and early growth \citep[e.g.][and references therein]{Woods18}. In particular, in order to match the observed masses at $z\approx6-7$, BH-seed models require extended periods of (possibly obscured) Eddington-limited,\footnote{The Eddington luminosity is defined as \mbox{$L_{Edd}=1.26\times10^{38}(\frac{M_{BH}}{M_\odot})$}} or even super-Eddington accretion, during which the structure and physics of the accretion may be different than at lower redshift, where QSOs are typically characterized by somewhat lower Eddington ratios \citep[e.g.][]{Shen12b}. This could produce a change of the $\alpha_{ox}-L_{UV}$ relation at high redshift.

Several works have compared the optical/UV continuum and emission-line properties (e.g. \citealt{Derosa14, Shen19}) of QSOs at $z>6$ and at lower redshifts, generally finding a lack of evident evolution. However, the fraction of weak-line QSOs (WLQs, i.e. objects with C IV and Ly$\alpha$+N~V rest-frame equivalent widths $REW<10,\angstrom$ and $REW<15\,\angstrom$, respectively; e.g. \citealt{Fan99}, \citealt{Diamond-Stanic09}) has been suggested to increase toward high redshift \citep[e.g.][]{Luo15, Banados16}, in spite of the color selection used for $z>6$ QSOs that may be biased against objects with weak $Ly\alpha$ lines \citep{Banados16}. Since WLQs are accreting preferentially with high Eddington ratios \citep[e.g.][]{Luo15,Marlar18}, the higher fraction of WLQs may indicate that the known QSOs at $z>6$ are generally accreting at higher Eddington ratios than at lower redshift, consistently with previous findings \citep[e.g.][]{Wu15}. \cite{Shen19} recently found an excess of weak-line QSOs (WLQs) at $z>5.7$ compared to lower redshifts. \cite{Meyer19} reported a strong increase of the typical blueshift of the C IV emission line in QSOs at $z\gtrsim6$, which can be linked again with the presence of a higher fraction of WLQs at high redshift (e.g. \citealt{Luo15}). About half of the WLQ population is found to emit significantly weaker X-ray radiation than the expectation based on the UV luminosity (e.g. \citealt{Ni18}), possibly linked to shielding by a geometrically thick inner accretion disk (e.g. \citealt{Luo15}), expected in the case of high Eddington-rate accretion.

X-ray observations can provide useful insights into the accretion physics in an independent way and on smaller scales than those probed by optical/UV emission. 
For instance, in addition to the $\alpha_{ox}$ parameter \citep[e.g.][]{Lusso17}, the intrinsic photon index ($\Gamma$) of the hard X-ray power-law continuum carries information about the coupling
between disk emission and the corona, and it is considered a proxy of the accretion rate. The relation between $\Gamma$ and the Eddington ratio has been established over a range
of redshifts for sizable samples of sources: steeper slopes correspond to higher implied Eddington
ratios \citep[e.g.][but see also \citealt{Trakhtenbrot17a}]{Shemmer08,Risaliti09,Brightman13,Fanali13}. 

Despite the large number of $z>6$ QSOs discovered to date, only $\approx15$ (i.e. $\approx8\%$ of the known population at these redshifts) are currently covered by sensitive pointed or serendipitous X-ray observations and only 11 are detected, severely limiting our ability to use X-rays to investigate the accretion physics and structure in QSOs in the early universe.
In this work, we present new \chandra\, observations for a sample of 10 QSOs at $z>6$. Along with archival data, we use these observations to constrain the X-ray properties of QSOs at $z>6$, derive the $\alpha_{ox}$ and $\Gamma$ parameters, and study possible dependencies upon redshift and luminosity. Our targets were selected to have virial estimates for BH masses from the Mg~II emission line, allowing us to include Eddington ratios in our analysis.

We adopt a flat cosmology with $H_0=67.7\,\mathrm{km\,s^{-1}}$ and $\Omega_m=0.307$ \citep{Planck16}.

\section{The sample of $z>6$ QSOs}
\subsection{Targets of new X-ray observations}

\begin{figure}
	\begin{center}
		\hbox{
			\includegraphics[width=90mm,keepaspectratio]{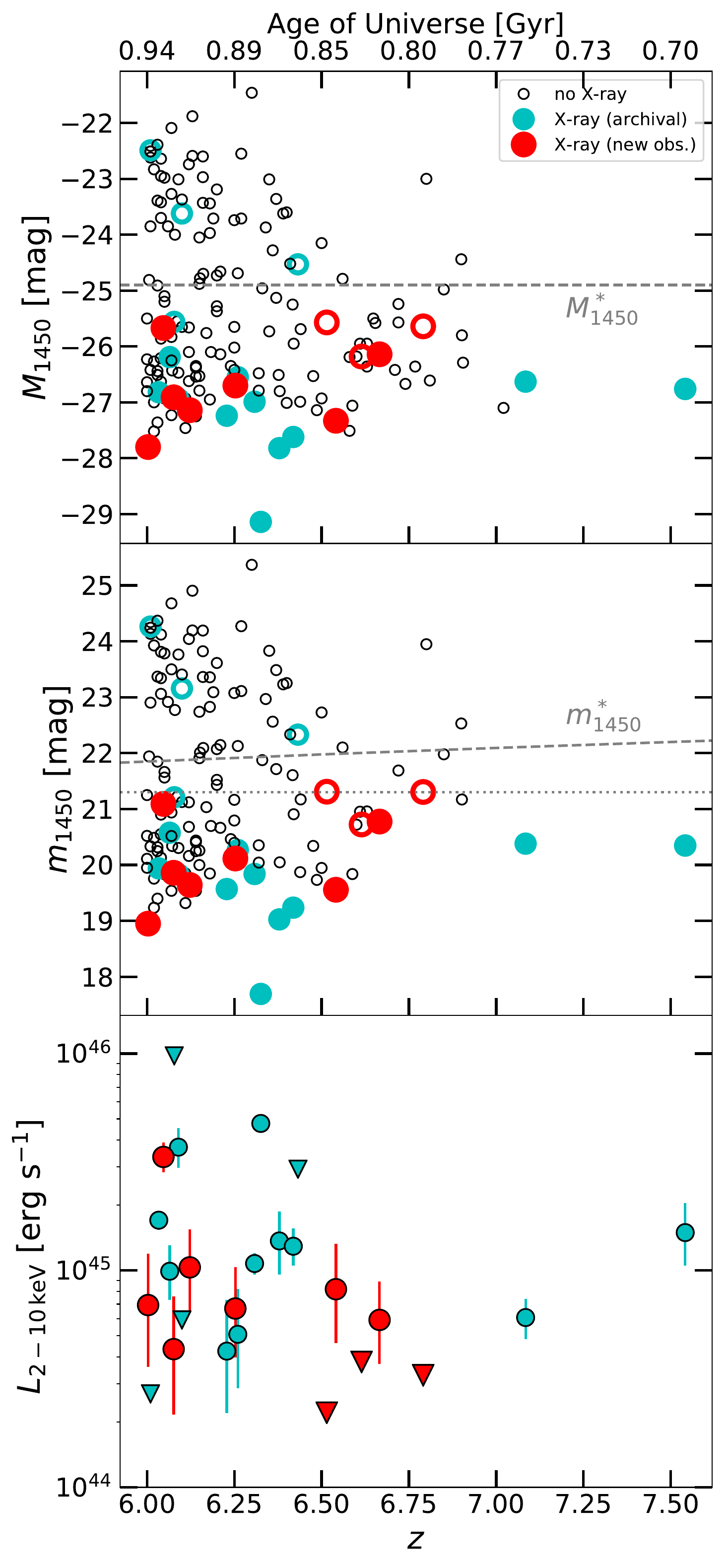} 
		}
	\end{center}
	\caption{\textit{Top and middle panels:} distribution of $M_{1450\angstrom}$ and $m_{1450\angstrom}$ as a function of redshift. Small black open circles are QSOs not covered by X-ray observations \citep{Banados16,Mazzucchelli17b,Reed17,Tang17,Wang17,Wang18a,Wang18b,Chehade18,Matsuoka18a,Matsuoka18b,YangJ18}. Cyan symbols are QSOs with archival X-ray data (see Tab.~\ref{Tab_sample_properties}). Red symbols are QSOs covered by the new \mbox{X-ray} observations presented in this work. Filled symbols are \mbox{X-ray} detected, open symbols are not detected. Dashed lines represent the break magnitudes of the QSO luminosity function \citep{Matsuoka18c}. The dotted line represents our magnitude selection. \textit{Bottom panel}: X-ray luminosity (derived as described in \S~\ref{data_analisys}) as a function of redshift. Symbols are the same as above, but upper limits on X-ray luminosity for undetected QSOs are shown as downward-pointing triangles.
		}	\label{Fig_mag_lum}
\end{figure}

We obtained \chandra\, observations of a sample of 10 type~1 QSOs at $z=6.0-6.8$ (Tabs.~\ref{Tab_sample_properties} and ~\ref{Tab_X-ray_obs}), with virial estimates of $M_{\mathrm{BH}}$ from near-IR spectroscopy (using the Mg~II line\footnote{Typical uncertainties for single-epoch mass estimates are $\gtrsim0.5$ dex (e.g. \citealt{Shen13} and references therein). In addition, the presence of spectral features (such as broad absorption lines) or weak emission lines can significantly affect the accuracy of the mass measurements.}; e.g. \citealt{Vestergaard09}). The targets were selected to be radio-quiet or, at most, radio-moderate QSOs (see \S~\ref{other_properties}). Five of them have absolute magnitudes $-26.2<M_{1450\angstrom}<-25.6$ (see red symbols in Fig.~\ref{Fig_mag_lum}), close to the break luminosity regime of the QSO luminosity function at $z\approx6$ (corresponding to $M_{1450\angstrom}\approx-24.9$; \citealt{Matsuoka18c}). This allows us to push the investigation of the X-ray emission of high-redshift QSOs down to a luminosity regime between typical SDSS QSOs \citep[e.g.][]{Paris18} and the fainter QSOs discovered by the SHELLQ survey \citep{Matsuoka16}. This region of the QSO $L-z$ parameter space has been probed poorly to date at X-ray wavelengths. In fact, the only four $M_{1450\angstrom}>-26$ QSOs at $z>6$ with previous X-ray data were serendipitously covered by X-ray observations (i.e. they were not targeted) and are not detected. Notably, with our new observations we more than triple the number of QSOs observed in X-rays at the highest redshifts ($z>6.5$). The distributions of the absolute and apparent magnitudes at rest-frame $1450\,\angstrom$ ($M_{1450\angstrom}$ and $m_{1450\angstrom}$, respectively) as a function of redshift are shown in Fig.~\ref{Fig_mag_lum} (top and middle panels), and are compared with known $z>6$ QSOs not observed in the X-rays.

\subsection{Other $z>6$ QSOs observed in X-rays}\label{arhive_sample}
\cite{Nanni17} studied the X-ray properties of all of the QSOs at $z>5.7$ previously covered by pointed or serendipitous X-ray observations, 14 of which are at $z>6$.
We include in our analysis these 14 $z>6$ QSOs. For these sources we used the magnitudes at $1450\,\angstrom$ provided by \cite{Banados16}. We also include ULASJ1342+0928, which was discovered after the \cite{Nanni17} work, and whose X-ray properties, magnitudes, and black-hole mass have been presented by \cite{Banados18a, Banados18b}. We thus include in our analysis a total of 15 $z>6$ QSOs with sensitive\footnote{We do not consider very shallow X-ray surveys, like the ROSAT All-Sky survey, which would provide only very loose upper limits on the X-ray fluxes of high-redshift QSOs.} archival observations in the X-ray band.

Seven of these QSOs were observed by \chandra\, only, three by \xmm\, only, four  by both \chandra\, and \xmm, and one by \textit{Swift}. 
Recently SDSSJ1030+0524 has been the target of a long \chandra\, imaging campaign (\mbox{$\approx480$}~ks, \citealt{Nanni18}), and was previously observed with both \chandra\, (with a shallow 8 ks observation; \citealt{Brandt02}) and \xmm\, (75 ks after background filtering; \citealt{Farrah04}). However, considering the long separation between the old and new observations ($\approx10-15$ years in the observed frame), and the hints for strong variation affecting its flux during this timespan, as discussed in \cite{Nanni18}, we limited our analysis to the deep 2017 \chandra\, dataset. Similarly, we consider only the $\approx80$ ks \chandra\, observation of SDSSJ1148+5251  \citep{Gallerani17}, and discarded a 2004 \xmm\, observation with a nominal exposure time of $\approx26$ ks, which is however almost completely affected by background flaring. As a result, for nine QSOs out of the 15 objects with archival observations we used only \chandra\, data, for 3 QSOs we used only \xmm \,data, for 2 QSOs we used data from both observatories, and for one we used \textit{Swift} data (see Tab.~\ref{Tab_X-ray_obs}). 

We searched the literature to retrieve black-hole mass estimates for these 15 QSOs (see Tab.~\ref{Tab_sample_properties}).
Since different authors used different calibrations to obtain estimates of black-hole masses, we recalibrate the values found in the literature to match the calibration of \cite{Vestergaard09}, as marked in Tab.~\ref{Tab_sample_properties}. We also modified luminosities and masses for our chosen cosmology. 
Furthermore, for consistency, we applied the same X-ray analysis (see \S~\ref{data_analisys}) to these archival observations.

\subsection{General properties of the sample}\label{other_properties}

The main physical properties of our sample are reported in Tab.~\ref{Tab_sample_properties}. For many of our targets, slightly different redshift values are reported in the literature, derived from the \mbox{Mg II} ($2799\,\angstrom$) and [C II] ($158\,\unit{\mu m}$) emission lines. When a \mbox{[C II]} measurement is available, we adopt it since the [C II] line is considered a better indicator of the systemic redshift than the Mg II line (e.g. \citealt{Decarli18}), which sometimes displays significant blueshifts in the observed wavelength \citep[e.g.][]{Plotkin15,Shen16}, possibly due to outflowing material in the broad emission-line region \citep[e.g. $\approx1700\,\mathrm{km\,s^{-1}}$ for SDSSJ0109$-$3047, corresponding to $\Delta z \approx0.04$;][]{Venemans16}. 

We computed the bolometric luminosities ($L_{bol}$) consistently for all our targets using the bolometric correction of \cite{Venemans16}, which was also used in \cite{Decarli18}: $\mathrm{log}(\frac{L_{bol}}{\mathrm{erg\,s^{-1}}})=4.553+0.911\times\mathrm{log}(\frac{\lambda L_{\lambda}(1450\angstrom)}{\mathrm{erg\,s^{-1}}})$. The typical uncertainty on $L_{bol}$ derived with this relation is $\sim7\%$. We thus provide homogeneously derived $L_{bol}$ rather than compiling values found in the literature, which are derived using different indicators of the bolometric luminosity (i.e. $L_{3000\angstrom}$ and $M_{1450\angstrom}$) and different bolometric corrections.

None of the QSOs included in our sample has been detected in the FIRST  \citep[which covers 16 of the 25 QSOs in our sample]{Becker95} or NVSS \citep[covering all of our $z>6$ QSOs]{Condon98} radio surveys. 
We report in Tab.~\ref{Tab_sample_properties} the radio-loudness parameter \mbox{$R=f_{\nu,5\mathrm{GHz}}/f_{\nu,4400 \angstrom}$} \citep{Kellerman89},  i.e. the ratio of the flux densities at rest-frame $5\,\mathrm{GHz}$ and $4400\,\angstrom$, or its upper limit, for the QSOs in our sample. 
$R$ values of QSOs included in the compilation of \cite{Banados15b} are taken from that work, including the only two QSOs detected at 1.4 GHz (with $R=0.7-0.8$).
For the remaining sources, we derived $f_{\nu,4400\angstrom}$ from $m_{1450\angstrom}$ (column 5 of Tab.~\ref{Tab_sample_properties}), assuming a power-law continuum with $\alpha=-0.3$, following \cite{Banados16}. 
Upper limits on the radio emission at $1.4$ GHz are derived as $3\times rms$ of the FIRST or NVSS surveys. For CFHQSJ0216$-$0455, we used the $rms$ of the VLA observations in the SXDS field \citep{Simpson06}. Finally, we estimated the upper limits on $f_{\nu,5\mathrm{GHz}}$ assuming a power-law spectrum with $\alpha=-0.75$.
Based on their upper limits on $R$, all of our sources are either radio-quiet ($R<10$) or at most radio-intermediate ($R<40$). We thus do not expect their X-ray emission to be significantly affected by a jet-linked contribution \citep[e.g.][]{Miller11}. \cite{Banados15b} reported a radio-loud QSO fraction of $\approx8\%$ among the $z\approx6$ population. The only three radio-loud QSOs at $z>6$ are not covered by X-ray observations and thus are not included in our sample.

It is difficult to establish firmly how many of the QSOs in the sample can be classified as WLQs, mainly because of the limited quality of the optical/UV spectra and spectral coverage. Beside the known WLQ SDSSJ0100+2802 \citep{Wu15}, other WLQ candidates are VIKJ0109$-$3047, ULASJ1342+0928, and SDSSJ2310+1855, all with $REW(\mathrm{C IV})\approx10-15$ (see Tab.~\ref{Tab_sample_properties} for the spectral references). However, several of the sources lack measurements of $REW(\mathrm{C IV})$. Furthermore, as reported in Tab.~\ref{Tab_sample_properties}, two QSOs are classified as broad absorption-line QSOs (BALQSOs), which usually show weak X-ray emission as well \citep[e.g.][]{Gallagher06, Gibson09a,Wu10,Luo14}.

\begin{table*}\tiny
	\caption{Physical properties of the $z>6$ QSOs with new or archival X-ray observations.}
	\begin{tabular}{cccccccccc} 
		\hline
		\multicolumn{1}{c}{{ ID }} &
		\multicolumn{1}{c}{{ RA}} &
		\multicolumn{1}{c}{{ DEC }} &
		\multicolumn{1}{c}{{ $z$}} &
		\multicolumn{1}{c}{{ $M_{1450\angstrom}$ ($m_{1450\angstrom}$)}} &
		\multicolumn{1}{c}{{ log$(\frac{L_{bol}}{L_\odot})$}} &
		\multicolumn{1}{c}{{ log$(\frac{M_{BH}}{M_\odot})$}} &
		\multicolumn{1}{c}{{ $\lambda_{Edd}$}} &
		\multicolumn{1}{c}{ Ref. (disc./$z$/$M_{BH}$)} &
		\multicolumn{1}{c}{ $R$} \\ 
		(1) & (2) & (3) & (4) & (5) & (6) & (7) & (8) & (9) & (10)\\
		\hline
		\multicolumn{9}{c}{{ New targets}} \\
		CFHQSJ0050+3445  & 00:50:06.67    & +34:45:21.65     & 6.253 (Mg II)         &  $-26.70$ (20.11)   & 13.45   & 9.41   & 0.34        &W10/W10/W10 & $<11.4$\\
		VIKJ0109$-$3047      & 01:09:53.13     & $-$30:47:26.31  & 6.7909 ([C II])          &  $-25.64$ (21.30)   & 13.06    & 9.12  & 0.27        & V13/V16/M17 & $<34.1$\\
		PSOJ036+03       &  02:26:01.87    & $+$03:02:59.42 & 6.541 ([C II])            &  $-27.33$ (19.55)   &  13.67   &9.48   & 0.48        & V15/B15/M17 & $<2.1$\\
		VIKJ0305$-$3150       &  03:05:16.92    & $-$31:50:55.9   &  6.6145 ([C II])         &  $-26.18$ (20.72)   &  13.26  & 8.95   & 0.63      &  V13/V16/M17& $<20.0$\\
		SDSSJ0842+1218       &  08:42:29.43 & +12:18:50.58      & 6.0763 ([C II])$^a$  &  $-$26.91 (19.86)$^a$   & 13.52   &  9.29 & 0.53       &  dR11/D18/dR11$^{*a}$& $<1.3$\\
		PSOJ167$-$13      &   11:10:33.98    &  $-$13:29:45.60 & 6.5148 ([C II])$^b$  &    $-$25.57 (21.25)  &  13.03 &  8.48  & 1.11       & V15/M17/M17& $<34.3$\\
		CFHQSJ1509$-$1749 & 15:09:41.78     &  $-$17:49:26.80  &  6.1225 ([C II])$^a$  & $-27.14$ (19.64)$^a$   & 13.61    &  9.47  & 0.42       &    W07/D18/W10$^a$& $<1.2$\\
		CFHQSJ1641+3755    & 16:41:21.73      &  +37:55:20.15     & 6.047 (Mg II)         &   $-25.67$ (21.09)   &  13.07  &   8.38  & 1.51       &   W07/W10/W10& $<10.5$\\
		PSOJ338+29       &  22:32:55.14    &  +29:30:32.31    & 6.666 ([C II])            &   $-26.14$ (20.78)  & 13.24   &  9.43  & 0.20        &  V15/M17/M17& $<21.0$\\
		SDSSJ2310+1855       &  23:10:38.89    &  +18:55:19.93      & 6.0031 ([C II])           &   $-27.80$ (18.95)  &  13.85  & 9.62 & 0.52       &  Wa13/Wa13/J16& $<3.9$\\
		\hline
		\multicolumn{9}{c}{{ QSOs with previous X-ray data}} \\
		SDSSJ0100+2802 & 01:00:13.02   & +28:02:25.92    & 6.3258 ([C II]) & $-29.14$ (17.69) & 14.33 & 10.03 & 0.62 &  Wu15/Wa16/Wu15$^*$& $<1.2$\\
		ATLASJ0142$-$3327 & 01:42:43.73 & $-$33:27:45.47   & 6.379 ([C II])$^a$ & $-27.82$ (19.02)$^a$ & 13.85 & --- & --- &  C15/D18/--- &$<4.2$\\
		CFHQSJ0210$-$0456   & 02:10:13.19  & $-04$:56:20.90 & 6.4323 ([C II])  &$-24.53$ (22.33) & 12.65 & 7.90 & 1.76 & W10/W13/W10& $<28.1$\\
		CFHQSJ0216$-$0455   & 02:16:27.81  & $-$04:55:34.10 & 6.01 (Ly $\alpha$)  &  $-22.49$ (24.27)  & 11.91    & ---    & ---      & W09/W09/---& $<23.1$\\		
		SDSSJ0303$-$0019      & 03:03:31.40  & $-$00:19:12.90 &6.078 (Mg II)   & $-25.56$ (21.21)  & 13.03   & 8.61 & 0.81 &  J08/K09/dR11$^*$& $<11.4$\\
		SDSSJ1030+0524         & 10:30:27.11   & +05:24:55.06    & 6.308 (Mg II) & $-26.99$ (19.84) & 13.55 & 9.21 & 0.68 &  F01/K07/dR11$^*$& $<1.5$\\
		SDSSJ1048+4637$^c$  &10:48:45.07    & +46:37:18.55    & 6.2284  (CO 6-5) & $-27.24$ (19.57) & 13.64 & 9.55 & 0.38 &   F03/Wa10/dR11$^*$ & $<0.5$\\
		ULASJ1120+0641          & 11:20:01.48   & +06:41:24.30   & 7.0842 ([C II]) & $-26.63$ (20.38) & 13.42 & 9.39 & 0.33 &  M11/V12/M17& $<0.7$\\
		SDSSJ1148+5251          & 11:48:16.65   & 52:51:50.39       & 6.4189  (CO 6-5) & $-27.62$ (19.24) & 13.78 & 9.71 & 0.36 & F03/Wa11/dR11$^*$& $0.7^{+0.2}_{-0.2}$\\
		SDSSJ1306+0356         & 13:06:08.27  & +03:56:26.36     &  6.0337 ([C II])$^a$           &  $-26.82$ (19.94)$^a$  & 13.49 & 9.30 &0.48 &     F01/D18/dR11$^{*a}$& $<1.5$\\
		ULASJ1342+0928         & 13:42:08.27  & +09:28:38.61     & 7.5413 ([C II]) & $-26.76$ (20.34) & 13.47 & 8.89 & 1.14 & B18a/V17/B18a& $<4.7$\\
		SDSSJ1602+4228         &  16:02:53.98 & +42:28:24.94     & 6.09 (Ly $\alpha$) &  $-26.94$ (19.83) & 13.53 & --- & ---  & F04/F04/---& $0.8^{+0.2}_{-0.2}$\\
		SDSSJ1623+3112          & 16:23:31.81   & +31:12:00.53     &  6.26 ([C II])  & $-26.55$ (20.27) & 13.39 & 9.15  & 0.54 & F04/Wa11/dR11$^*$ & $<2.3$\\
		SDSSJ1630+4012          &  16:30:33.90 &  +40:12:09.69    &  6.065 (Mg II)    & $-26.19$ (20.58) & 13.26   & 8.96  &    0.62& F03/I04/dR11$^*$& $<2.2$\\
		HSCJ2216$-$0016$^c$ & 22:16:44.47  & $-$00:16:50.10   &  6.10 (Ly $\alpha$) & $-23.62$ (23.16) & 12.32  & --- & --- &    M16/M16/---& $<40.9$\\
		\hline
	\end{tabular} \label{Tab_sample_properties}\\
	\small (1): QSO ID. (2) and (3): RA and DEC (J2000) from \cite{Banados16,Banados18a}. (4): Redshift and emission line from which it is derived. In cases of different values derived from different emission lines for the same source, we preferred the redshift derived from the [C II] line rather than the Mg II line, as discussed in \S~\ref{other_properties}. (5): Absolute and apparent magnitude at $1450\,\angstrom$. Note that $M_{1450\angstrom}$ can vary by up to $\approx0.3$ mag among different papers, depending on the prescription used to compute it (e.g. \citealt{Omont13} vs. \citealt{Banados16} for CFHQSJ1641+3755). We consistently assumed the values reported by \citet{Banados16} and \citet{Mazzucchelli17b}, which used the same prescription, for all our sources. (6): Bolometric luminosity estimated from $M_{1450\angstrom}$, using the bolometric correction of \citet{Venemans16}.  (7): Virial black-hole mass estimated from the Mg II emission line. Note that \citet{Trakhtenbrot17b} used different calibrations for the black-hole masses of several QSOs included in our sample, typically resulting in larger values (up to $\approx0.2-0.3$ dex).  (8): Eddington ratio: $\lambda_{Edd}=L_{bol}/L_{Edd}$. (9): Reference for the QSO discovery, adopted redshift, and black-hole mass. B15: \citet{Banados15}; B18a: \citet{Banados18a}; C15: \citet{Carnall15}; dR11: \citet{deRosa11}; D18:  \citet{Decarli18}; F01:  \citet{Fan01};  F03:  \citet{Fan03}; F04:  \citet{Fan04}; I04: \citet{Iwamuro04}; J08: \citet{Jiang08}; J16: \citet{Jiang16}; K07: \citet{Kurk07}; K09: \citet{Kurk09}; M11: \citet{Mortlock11}; M16: \citet{Matsuoka16}; M17: \citet{Mazzucchelli17b}; V12: \citet{Venemans12}; V13: \citet{Venemans13}; V15: \citet{Venemans15a}; V16: \citet{Venemans16}; V17: \citet{Venemans17}; W07: \citet{Willott07}; W09: \citet{Willott09}; W10: \citet{Willott10b}; W13: \citet{Willott13}; Wa10: \citet{Wang10}; Wa11: \citet{Wang11}; Wa13: \citet{Wang13}; Wa16: \citet{Wang16}; Wu15: \citet{Wu15}. (10): radio-loudness parameter (see \S~\ref{other_properties})\\
	\textit{Notes}: $^*$ For these QSOs, black-hole masses have been modified according to the \citet{Vestergaard09} calibration, to be consistent with the other QSOs. In these cases, the references indicate the papers from which FWHM(Mg II) and $L_{3000\angstrom}$ are collected.  $^a$ For these sources we updated magnitudes and black-hole masses according to the new \mbox{[C II]-based} redshifts provided by \citet{Decarli18}. These values differ negligibly from those derived assuming previous redshifts based on Mg II or Ly$\alpha$ emission lines. 
	$^b$ \cite{Willott17} independently reported a slightly different value ($z=6.5157$) from [C II]. $^c$ Broad absorption-line QSOs (see \citealt{Fan04,Matsuoka16}) . 
\end{table*}

\section{Data analysis}\label{data_analisys}
\subsection{X-ray data reduction}\label{data_reduction}
Tab.~\ref{Tab_X-ray_obs} summarizes the basic information about the X-ray observations of our new targets and archival sources. 
We reprocessed the \chandra\, observations with the \textit{chandra\_repro} script in CIAO 4.10,\footnote{\url{http://cxc.harvard.edu/ciao/}} using CALDB v4.8.1,\footnote{\url{http://cxc.harvard.edu/caldb/}} setting the option \textit{check\_vf\_pha=yes} in the case of observations taken in very faint mode. 
We created exposure maps with the \textit{fluximage } script. Spectra, response matrices, and ancillary files for sources and associated background were extracted using the \textit{specextract } tool. 

SDSSJ1030+0524 has been observed with \mbox{ACIS-I} in ten individual pointings over five months, for a total of $\approx480$ ks \citep[see Tab.~\ref{Tab_X-ray_obs} and][]{Nanni18}. Similarly,  VIKJ0109--3047, SDSSJ1306+0356, ULASJ1342+0928, and CFHQSJ1641+3755 have been targeted with two \chandra\, observations, for a total of $\approx65$, $126$, $45$, and $54$ ks, respectively  \citep[see Tab.~\ref{Tab_X-ray_obs} and][]{Banados18b}. For these sources, we checked for astrometry issues and merged the individual observations with the \textit{reproject\_obs} tool, and derived merged images and exposure maps. In doing this, we effectively combine the different pointings into a single, longer exposure. Spectra, response matrices, and ancillary files extracted from the single pointings were added using the \textit{mathpha}, \textit{addrmf}, and \textit{addarf} HEASOFT tools\footnote{\url{https://heasarc.gsfc.nasa.gov/docs/software/heasoft/}}, respectively, weighting by the individual exposure times.

\xmm\, observations have been processed with SAS v16.1.0., following the standard procedure\footnote{\url{https://www.cosmos.esa.int/web/xmm-newton/sas-threads}} and filtering for periods of high background levels imposing count-rate thresholds of $<0.4$ and $<0.35\,\mathrm{cts\,s^{-1}}$ in the $10<E<12$ keV and $E>10$ keV bands for the EPIC/PN and EPIC/MOS cameras, respectively. We created images and exposure maps, and extracted  spectra, response matrices, and ancillary files using the \textit{evselect},\textit{eexpmap}, \textit{backscale}, \textit{rmfgen}, and \textit{arfgen} tools.

In the case of sources targeted by multiple \xmm\, pointings (i.e. CFHQSJ0210$-$0456 and ULASJ1120+0641; see Tab.~\ref{Tab_X-ray_obs}), we merged the different datasets for each EPIC camera with the \textit{merge} tool, and, similarly to what we did for \chandra\, sources, we added the spectra extracted from each observation with the \textit{epicspeccombine} tool. We also averaged the response matrices and ancillary files with the \textit{addrmf} and \textit{addarf} tools, weighting by the exposure times of the individual observations.\footnote{Note that \textit{epicspeccombine} returns as output a summed spectrum with exposure time set to the average value of the two input spectra, and the sum of the two input ancillary files. This is the equivalent of observing the source for half of the total time with a fictional camera with twice the sensitivity of the actual camera. By changing with \textit{dmhedit} the exposure time keyword of the output summed spectrum to the summed exposure time of the two input spectra, and by computing the weighted average of the response matrices and ancillary files with \textit{addrmf} and \textit{addarf}, we return to the case in which the source is observed by the actual camera for a longer exposure time. The two cases are equivalent when spectra and ancillary files are used together (e.g. performing spectral analysis with XSPEC). However, in \S~\ref{photometry} we will use the ancillary files alone to compute the count-rate to flux conversion factors. In such a case, the use of the summed ancillary files obtained as output of \textit{epicspeccombine} would not be correct.} We then used the merged images to compute source photometry (see \S~\ref{detection}). Since these sources were placed at similar off-axis angles in the different 
pointings, by merging the observations for each camera we effectively combine them into single and longer observations. However, we keep the different cameras separated, as the responses are significantly different. We then combined the scientific results, as described in \S~\ref{photometry}. 

We reduced \textit{Swift}-XRT data for ATLASJ0142$-$3327 as in \cite{Nanni17}, using the standard software (HEADAS v. 6.18)\footnote{\url{https://heasarc.gsfc.nasa.gov/docs/software/lheasoft/}} and procedures.\footnote{\url{https://swift.gsfc.nasa.gov/analysis/}} An ancillary file has been extracted with the \textit{xrtmkarf} tool.

\begin{table}\scriptsize
	\caption{Summary of our new \chandra\, and archival X-ray observations of the sample of $z>6$ QSOs.}
	\begin{tabular}{ccccccccc} 
		\hline
		\multicolumn{1}{c}{{ ID }} &
		\multicolumn{1}{c}{{ OBSID}} &
		\multicolumn{1}{c}{{ Date }} &
		\multicolumn{1}{c}{{ $T_{exp}$ [ks]}} \\ 
		\hline
		\multicolumn{4}{c}{{ New observations }} \\
		CFHQSJ0050+3445$^\mathrm{C}$   & 20393 &2017-09-25 &33.5\\
		VIKJ0109$-$3047$^\mathrm{C}$      & 20398 & 2019-05-07 & 37.0\\
		"$^\mathrm{C}$                        & 22214    & 2019-05-10  & 29.5 \\
		PSOJ036+03$^\mathrm{C}$      & 20390 & 2018-10-09& 25.9 \\
		VIKJ0305$-$3150$^\mathrm{C}$       &20394 &2018-05-11 & 49.9 \\
		SDSSJ0842+1218$^\mathrm{C}$       &  20392 & 2018-01-01 &28.7 \\
		PSOJ167$-$13$^\mathrm{C}$      &  20397 & 2018-02-20 &59.3 \\
		CFHQSJ1509$-$1749$^\mathrm{C}$ & 20391 &2018-06-06 & 26.8\\
		CFHQSJ1641+3755$^\mathrm{C}$    & 20396 & 2018-11-15 & 20.8\\
		"$^\mathrm{C}$                       & 21961 & 2018-11-17 & 33.5\\ 
		PSOJ338+29$^\mathrm{C}$       & 20395 & 2018-01-30 & 54.2\\
		SDSSJ2310+1855$^\mathrm{C}$      &  20398 & 2017-09-30 & 17.9\\
		\hline
		\multicolumn{4}{c}{{ Archival observations }} \\	
		SDSSJ0100+2802$^\mathrm{C}$ & 17087 & 2015-10-16 & 14.8 \\
		" $^\mathrm{X}$ & 0790180701 & 2016-06-29 & 44.9/60.7/60.4 \\
		ATLASJ0142$-$3327$^\mathrm{S}$ & 00290624001 & 2007-09-11  & 20.9\\
		CFHQSJ0210$-$0456$^\mathrm{X}$ &  0677630133 & 2012-07-10 &  8.0/10.6/10.6\\
		" $^\mathrm{X}$ & 0677640133 & 2012-01-12 & 8.4/10.6/10.5\\
		CFHQSJ0216$-$0455$^\mathrm{X}$ & 0112370601 & 2002-08-12 & 29.4/37.9/37.9\\
		SDSSJ0303$-$0019$^\mathrm{C}$&13349& 2011-11-27&1.5\\
		SDSSJ1030+0524$^\mathrm{C}$ & 18185 & 2017-01-17 & 46.3 \\
		"$^\mathrm{C}$                           &18186  &  2017-01-25 &34.6   \\
		"$^\mathrm{C}$                           & 18187 &2017-03-22   &  40.4 \\
		"$^\mathrm{C}$                           & 19926 & 2017-05-25  & 49.4  \\
		"$^\mathrm{C}$                           & 19987 &2017-01-18   &126.4   \\
		"$^\mathrm{C}$                           & 19994 & 2017-01-27  & 32.7  \\
		"$^\mathrm{C}$                           & 19995 &  2017-01-27 & 26.7  \\
		"$^\mathrm{C}$                           & 20045 &  2017-03-24 & 61.3  \\
		"$^\mathrm{C}$                           & 20046 & 2017-03-26  &  36.6 \\
		"$^\mathrm{C}$                           & 20081 & 2017-05-27  & 24.9  \\
		SDSSJ1048+4637$^\mathrm{C}$ & 5608 & 2005-01-10&15.0\\
		ULASJ1120+0641$^\mathrm{C}$ & 13203 & 2011-02-04 & 15.8\\
		"$^\mathrm{X}$ & 0693990101 & 2012-05-23 & 24.1/46.5/45.8\\
		"$^\mathrm{X}$ & 0693990201 &2012-06-18 & 71.9/108.0/108.1\\
		"$^\mathrm{X}$ & 0693990301 & 2012-06-20& 56.4/83.6/84.1\\
		SDSSJ1148+5251$^\mathrm{C}$  & 17127 & 2015-09-02& 77.8 \\
		SDSSJ1306+0356$^\mathrm{C}$ & 3358 & 2002-01-29 & 		8.2 \\
		"$^\mathrm{C}$                         & 3966 & 	2003-11-29 & 118.2 \\
		ULASJ1342+0928$^\mathrm{C}$ & 20124 & 2017-12-15 &  24.7\\
		"$^\mathrm{C}$                         & 20887 & 2017-12-17 & 20.4 \\
		SDSSJ1602+4228$^\mathrm{C}$&5609 & 2005-10-29 & 13.2 \\
		SDSSJ1623+3112$^\mathrm{C}$ & 5607 & 2004-12-29 & 17.2 \\
		SDSSJ1630+4012$^\mathrm{C}$&	5618 & 2005-11-04 & 27.4 \\	
		HSCJ2216$-$0016$^\mathrm{X}$ & 0673000145 & 2011-12-08 & 3.7/4.2/4.2 \\
		\hline
	\end{tabular} \label{Tab_X-ray_obs}\\
	\textit{Notes:} $^\mathrm{C}$ source observed with \chandra. $^\mathrm{X}$ source observed with \xmm. Exposure times are filtered for background flaring and correspond to the PN, MOS1, and MOS2 cameras, respectively.   $^\mathrm{S}$ source observed with \textit{Swift}.
\end{table}

\subsection{Detection procedure}\label{detection}

For \chandra\, observations, we used circular source extraction regions centered on the optical positions of the targets and with radii of 2 arcsec, to account for X-ray and optical positional uncertainties, and any possible small X-ray-to-optical offset. This region size encompasses $\approx100\%$ and $\approx90\%$ of the \chandra\, PSF at $E=1.5$ and 6.4 keV, respectively, for an on-axis position.  The background levels are evaluated in local annular regions centred on the targets, with inner and outer radii of 4 and 24 arcsec, respectively, free of contaminating sources. All the sources in our sample covered by \chandra\, observations were observed on axis, except for SDSSJ0303$-0019$, which is observed at an off-axis angle of $\approx4.8$ arcmin.

For \xmm\, observations, we used circular source extraction regions centered on the optical positions of the targets and with radii of 10--30 arcsec (corresponding to $\approx50-80\%$ of the PSF), depending on the off-axis angle of the source ($0-6$ arcmin) and the presence of nearby detected objects that could contaminate the photometry. Circular background extraction regions are placed at nearby locations free of evident detected sources and have radii of 60--80 arcsec. 
For the \textit{Swift}-XRT observation of ATLASJ0142$-$3327 we computed the source photometry in a circular region with radius 10 arcsec, which equates to $\approx$ 50\% of the PSF \citep{Moretti05}, and the background photometry in a nearby circular region with radius $\approx72$ arcsec.

We ran the detection procedure in three energy bands ($0.5-2$, $2-7$, and $0.5-7$ keV, which we refer to as the soft, hard, and full bands, respectively) separately for every available instrument (ACIS, EPIC/PN, EPIC/MOS1, EPIC/MOS2, and XRT). Different images of one object taken with the same instrument were merged, as described in \S~\ref{data_reduction}.
We computed the detection significance in each energy band using the binomial no-source probability \mbox{\citep{Weisskopf07,Broos07}}
\begin{equation}
P_B(X\geq S)= \sum_{X=S}^{N}\frac{N!}{X!(N-X)!}p^X(1-p)^{N-X},
\end{equation}
where $S$ is the total number of counts in the source region in the considered energy band, $B$ is the total number of counts in the background region, \mbox{$N=S+B$}, and $p=1/(1+BACKSCAL)$, with $BACKSCAL$ being the ratio of the background and source region areas. For sources observed by multiple instruments, we consider  the quantity \hbox{$P^{TOT}_B=\prod_{i}P^i_B$} as the final binomial no-source probability in one energy band, where the product is performed over all the instruments used to observe a source. 
We consider a source to be detected if $(1-P_B)>0.99$.  
Out of the 111 analyzed images (25 objects in the three energy bands, some of which were observed by different instruments, see Tab.~\ref{Tab_photometry}), we expect $111\times P_B\approx1$ false detection with the adopted significance threshold. 

\begin{figure*}
	\includegraphics[width=85mm,keepaspectratio]{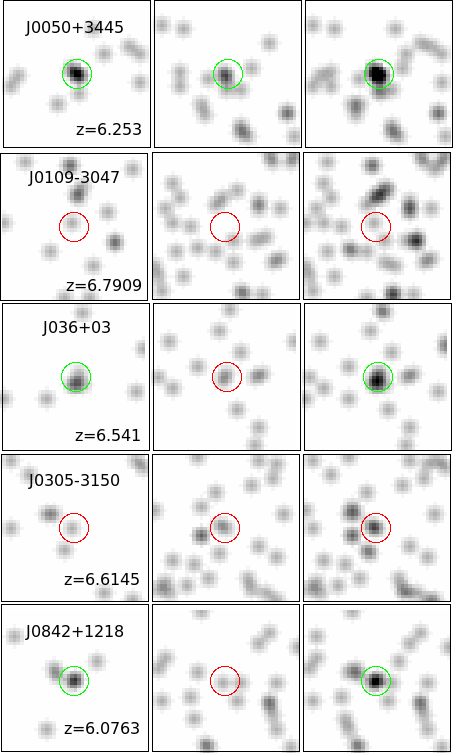} 
	\includegraphics[width=85mm,keepaspectratio]{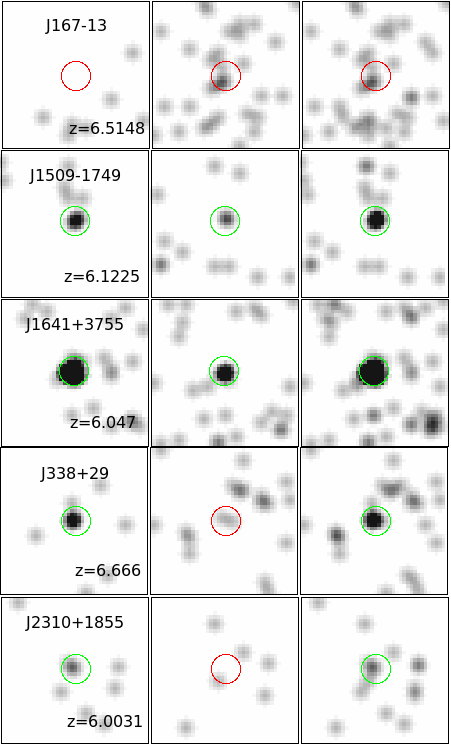} 
	\caption{Smoothed \chandra\, images ($40\times40$ pixels; i.e. $\approx 20''\times20''$) of our ten new targets (rows, as annotated) in the soft (first column), hard (second column), and full (third column) band. Circles represent the source extraction regions ($R=2$ arcsec) centred on the optical positions of the targets, and  used to compute source photometry. Green and red circles are used for detected and undetected sources, respectively.	}
	\label{Fig_cutouts1}
\end{figure*}

Fig.~\ref{Fig_cutouts1} displays the X-ray images of our new targets in the three energy bands (see \citealt{Nanni17} and \citealt{Banados18b} for similar images for the archival sources). Detected and undetected sources are identified with green and red circles, respectively. 
Three of our 10 observed targets are detected in all of the three considered bands, four QSOs are detected in the soft and full bands only, and three are not detected in any band. 
Reasonably different sizes for the source and background extraction regions do not affect these results.

\subsection{Photometry, fluxes, and luminosities}\label{photometry}

We computed the net counts and associated uncertainties (or upper limits in the case of non-detections) by deriving the probability distribution function of net counts with the method of \citet[see their Appendix A3]{Weisskopf07}, which correctly accounts for the Poisson nature of both source and background counts. For sources detected by an instrument in one energy band, we report in Tab.~\ref{Tab_photometry} the nominal value of the net counts, corresponding to the peak of the probability distribution, and the errors corresponding to the narrowest 68\% confidence interval. For undetected sources we report the  upper limit corresponding to the 90\% confidence interval. These values are not corrected for the fraction of PSF excluded in the extraction regions.

We used the probability distribution functions of the net counts in the soft and hard bands to constrain the hardness ratio $HR=(H-S)/(H+S)$, where $S$ and $H$ are the observed net counts in the soft and hard bands, respectively: we randomly picked a pair of values following such functions and computed $HR$. Repeating the procedure 10000 times, we constructed the probability distribution function of $HR$, and computed the $68\%$ confidence interval, or $90\%$ upper limit in the case of sources undetected in the hard band (Tab.~\ref{Tab_photometry}). We found no significantly different hardness-ratio values using the Bayesian Estimation of Hardness Ratios (BEHR) code \citep{Park06}.
The last column of Tab.~\ref{Tab_photometry} reports the effective photon indices corresponding to the $HR$ values, computed assuming a power law model and Galactic absorption \citep{Kalberla05}, and accounting for the effective area of each instrument at the time of each observation and at the position of each source on the detector.

The probability distribution functions of X-ray flux in the three energy bands have been derived from the net count-rate probability distribution function assuming a power-law spectrum with $\Gamma=2.0$ (typical of luminous QSOs, e.g. \citealt{Shemmer06b}, \citealt{Nanni17}, see also \S~\ref{spec_analysis}), accounting for Galactic absorption \citep{Kalberla05} and using the response matrices and ancillary files extracted at the position of each target. All of the ancillary files are corrected for the fraction of the PSF not included in the extraction regions. Thus, fluxes and derived quantities are corrected for PSF effects. Tab.~\ref{Tab_flux} reports the fluxes corresponding to the peak of the probability distribution functions, and the uncertainties corresponding to the narrowest interval containing 68\% of the total probability for sources detected in an energy band. For undetected objects we report the upper limit corresponding to the 90\% probability.

For QSOs observed by different instruments, we derived the flux probability distribution function for each instrument, multiplied them together and then renormalized the result to obtain the average distribution. This was used to compute the nominal fluxes and uncertainties. Deep observations produce narrower probability distribution functions than shallower pointings, and thus dominate the averaged final distribution.
This averaging procedure works if a source did not vary strongly between the different observations; otherwise, the flux probability distribution functions for the individual instruments do not overlap and their product is null. There is no such case in our sample. We note, however, that for some objects observed by multiple instruments several months apart (e.g. SDSSJ0100+2802 and ULASJ1120+0641), the flux probability distribution functions of the individual instruments are slightly shifted, although they still largely overlap. While this shift can be simply explained by statistical fluctuations of the measured counts, we cannot exclude some level of source variability. In this case our results would correspond to fluxes averaged over the different observed states. Note that \cite{Shemmer17} report no significant evolution of QSO X-ray variability amplitude with redshift, at least up to $z\approx4.3$.

Luminosities in the rest-frame $2-10$ keV band (Tab.~\ref{Tab_flux}) and monochromatic luminosities at 2 keV have been computed from the unabsorbed (i.e. corrected for Galactic absorption) fluxes in the soft band, assuming again $\Gamma=2.0$. Fig.~\ref{Fig_mag_lum} (bottom panel) presents the distribution of X-ray luminosity versus redshift for $z\ge6$ QSOs. A short extrapolation is needed in the X-ray luminosity calculation, since the emission at rest-frame 2 keV is redshifted below 0.5 keV at $z>6$, and is thus not directly probed by X-ray observations.

\begin{table*}\addtolength{\tabcolsep}{-2pt}    
	\centering
	\caption{Observed X-ray photometry and hardness ratios. Net counts and associated uncertainties are computed by deriving the probability distribution function of net counts with the method of \citet{Weisskopf07}.} 
	\begin{tabular}{cccccccccccccccc} 
		\hline
		\multicolumn{1}{c}{{ ID }} &
		\multicolumn{3}{c}{{ Net counts}} &
		\multicolumn{1}{c}{{ HR}} &
		\multicolumn{1}{c}{{ $\Gamma_{eff}$}}	\\ 
		\cline{2-4}
		\multicolumn{1}{c}{{ }} &
		\multicolumn{1}{c}{SB} &
		\multicolumn{1}{c}{{HB }} &
		\multicolumn{1}{c}{{ FB}} &
		\multicolumn{1}{c}{} &
		\multicolumn{1}{c}{}		\\	
		\hline
		\multicolumn{5}{c}{{ New observations }} \\
		CFHQSJ0050+3445&$ 4.8_{-1.9}^{+2.6}$  & $ 2.7_{-1.4}^{+2.1}$  &$7.4 _{-2.5}^{+3.2}$ & $ -0.26_{-0.39}^{+0.26}$   & $1.68_{-0.28}^{+1.10}$\\
		VIKJ0109$-$3047   & $<3.5$& $<2.3$& $<3.1$&  -- &--   \\
		PSOJ036+03    & $ 3.9_{-1.7}^{+2.4}$ &$<4.9$ & $ 5.5_{-2.1}^{+2.8}$ & $<0.02$ & $>1.12$  \\
		VIKJ0305$-$3150    & $<3.6$&$<4.7$ &$<5.7$ &--   &-- \\
		SDSSJ0842+1218    & $ 2.8_{-1.4}^{+2.1}$ & $<3.5$&$ 3.3_{-1.7}^{+2.4}$  & $ <0.17$ & $>0.77$ \\
		PSOJ167$-$13  &$<2.3$ &$<7.2$ &$ <6.8$  &--  & --\\
		CFHQSJ1509$-$1749 & $5.7_{-2.1}^{+2.8}$& $2.6_{-1.4}^{+2.1}$& $8.4_{-2.7}^{+3.4}$ & $-0.33_{-0.35}^{+0.26}$ & $1.94_{-0.60}^{+1.06}$\\
		CFHQSJ1641+3755  & $39.5_{-6.0}^{+6.6}$& $8.3_{-2.7}^{+3.4}$&$47.8_{-6.7}^{+7.3}$&  $-0.65_{-0.15}^{+0.08}$ & $2.15_{-0.19}^{+0.49}$\\
		PSOJ338+29     & $ 5.6_{-2.1}^{+2.8}$ & $<4.7$& $ 6.9_{-2.5}^{+3.2}$ & $ <-0.06$  & $>1.30$\\
		SDSSJ2310+1855    & $ 2.9_{-1.4}^{+2.1}$ & $<3.7$&$ 3.7_{-1.7}^{+2.4}$  & $<0.18$ & $>0.74$  \\
		\hline
		\multicolumn{5}{c}{{ Archival observations }} \\
		SDSSJ0100+2802 (\chandra) &  $12.8_{-3.3}^{+4.0}$ & $<5.1$ & $14.6_{-3.6}^{+4.2}$ & $<-0.42$ &$>1.88$  \\
		" (PN) &  $149.5_{-13.6}^{+14.3}$ & $31.2_{-8.4}^{+9.0}$ & $180.7_{-16.1}^{+16.8}$ &  $-0.66_{-0.11}^{+0.07}$ & $2.18_{-0.20}^{+0.38}$\\		
		" (MOS1) &  $74.6_{-9.1}^{+9.7}$ & $<11.3$ & $78.6_{-10.0}^{+10.6}$ & $<-0.72$ & $>2.68$\\		
		" (MOS2) &  $52.3_{-7.8}^{+8.4}$ & $11.4_{-4.5}^{+5.2}$ & $64.0_{-9.1}^{+9.8}$ &  $-0.64_{-0.16}^{+0.11}$ & $2.44_{-0.33}^{+0.62}$\\		
		ATLASJ0142-3327 (\textit{Swift}) & $11.0_{-3.3}^{+4.0}$ &	$<4.2$ &$11.5_{-3.5}^{+4.3}$ &$<-0.41$ & $>1.68$ \\
		CFHQSJ0210-0456 (PN) & $<14.2$ &$<18.4$ & $<12.3$ &-- & --\\
		" (MOS1) & $<11.2$ &$<4.5$ & $<7.1$ &   --  & -- \\
		" (MOS2) & $<17.3$ &$<14.2$ & $<13.5$ &  --  & -- \\	 			
		CFHQSJ0216$-$0455 (PN)& $<9.3$ & $<4.0$ & $<7.2$ & -- & --\\
		" (MOS1)              & $<8.5$ & $<7.0$  & $<11.5$ &  --  & --\\
		" (MOS2)              & $<3.4$ & $<3.4$  & $<3.7$ &  -- & -- \\		
		SDSSJ0303-0019(\chandra)& $ <2.3$ & $<3.9$ &$<3.9$ & -- & --\\
		SDSSJ1030+0524(\chandra) & $78.2_{-8.6}^{+9.2.}$  &$46.4_{-6.7}^{+7.4}$ & $124.6_{-11.0}^{+11.7}$ &$-0.26_{-0.11}^{+0.07}$& $1.91_{-0.17}^{+0.25}$ \\
		SDSSJ1048+4637(\chandra) & 	$ 2.9_{-1.4}^{+2.1}$& $<2.3$ & $ 2.8_{-1.4}^{+2.1}$ &$<-0.02$ & $>0.66$ \\
		ULASJ1120+0641 (\chandra) &   $3.9_{-1.7}^{+2.4}$ & $<5.1$ & $5.7_{-2.1}^{+2.8}$ &  $<0.19$ & $>0.39$\\
		" (PN)    & $21.2_{-8.3}^{+8.9}$& $<10.7$& $<32.3$& $<-0.21$& $>1.39$\\
		" (MOS1)    & $14.9_{-6.4}^{+7.0}$& $<6.2$& $<19.2$&$<-0.24$ & $>1.41$\\
		" (MOS2)    & $17.8_{-6.3}^{+7.0}$& $<24.2$& $32.3_{-9.1}^{+9.7}$& $<0.25$ & $>0.47$\\		
		SDSSJ1148+5251(\chandra) &  $26.5_{-4.9}^{+5.5}$ &$10.2_{-3.0}^{+3.7}$  &  $36.7_{-5.8}^{+6.5}$&  $-0.44_{-0.18}^{+0.12}$ & $1.87_{-0.29}^{+0.50}$\\
		SDSSJ1306+0356(\chandra) &		$ 105.0_{-9.9}^{+10.6}$	&$ 28.0_{-5.0}^{+5.7}$ & $ 133.1_{-5.0}^{+5.7}$ & $-0.57_{-0.10}^{+0.05}$ & $1.78_{-0.14}^{+0.28}$\\
		ULASJ1342+0928(\chandra) & $9.7_{-2.9}^{+3.5}$ &$4.4_{-1.9}^{+2.7}$ &$14.1_{-3.6}^{+4.2}$ &$-0.36_{-0.29}^{+0.20}$ & $1.88_{-0.44}^{+0.86}$\\
		SDSSJ1602+4228(\chandra) & $ 22.9_{-4.5}^{+5.1}$ & $ 3.7_{-1.7}^{+2.4}$ & $ 25.6_{-4.9}^{+5.5}$ & $-0.70_{-0.14}^{+0.13}$  &$2.21_{-0.39}^{+0.63}$ \\
		SDSSJ1623+3112(\chandra) & $ 3.9_{-1.7}^{+2.4}$ &  $ 2.9_{-1.4}^{+2.1}$ & $ 6.8_{-2.3}^{+3.0}$ & $-0.14_{-0.36}^{+0.31}$ & $0.89_{-0.58}^{+0.74}$\\
		SDSSJ1630+4012(\chandra) & $ 12.7_{-3.3}^{+4.0}$ & $ 4.8_{-1.9}^{+2.6}$	& $ 17.5_{-3.9}^{+4.6}$	& $-0.43_{-0.24}^{+0.18}$ & $1.47_{-0.07}^{+0.63}$\\
		HSCJ2216$-$0016 (PN)& $<4.7$ & $<3.9$ & $<5.5$& --& --\\
		" (MOS1)& $<2.3$ & $<2.3$ & $<2.3$& --& --\\
		" (MOS2 & $<2.3$ &$<2.3$  & $<2.3$& --& --\\
		\hline
	\end{tabular} \label{Tab_photometry}\\
\end{table*}

\begin{table*}
	\centering
	\caption{X-ray fluxes, luminosities, and derived properties for our \chandra\, and archival sample of $z>6$ QSOs. Errors account for the uncertainties on the net counts only.}
	\begin{tabular}{cccccccc} 
		\hline
		\multicolumn{1}{c}{{ ID }} &
		\multicolumn{3}{c}{{ $F$}} &
		\multicolumn{1}{c}{{ $L_{2-10\mathrm{keV}}$}} &
		\multicolumn{1}{c}{ $\alpha_{ox}$} &
		\multicolumn{1}{c}{ $\Delta\alpha_{ox}$} \\ 
		\multicolumn{1}{c}{} &
		\multicolumn{3}{c}{{ [$10^{-15}\,\funit$}]} &
		\multicolumn{1}{c}{{ [$10^{44}\,\lunit$]}} &
		\multicolumn{1}{c}{ } &
		\multicolumn{1}{c}{ } \\ 	
		\cline{2-4}
		\multicolumn{1}{c}{{ }} &
		\multicolumn{1}{c}{SB} &
		\multicolumn{1}{c}{{HB }} &
		\multicolumn{1}{c}{{ FB}} &
		\multicolumn{1}{c}{} &
		\multicolumn{1}{c}{} &
		\multicolumn{1}{c}{ } &
		\multicolumn{1}{c}{ } \\ 		
		\hline
		\multicolumn{7}{c}{{ New observations }} \\
		CFHQSJ0050+3445&$1.07_{-0.43}^{+0.59}$ & $1.48_{-0.80}^{+1.19}$&$2.46_{-0.83}^{+1.05}$ & $6.68_{-2.70}^{+3.67}$ &$-1.71_{-0.09}^{+0.07}$ & $-0.02_{-0.09}^{+0.07}$ \\
		VIKJ0109$-$3047   &$<0.47$ &$<0.63$ & $<0.56$& $<3.29$&$<-1.67$ & $<-0.04$     \\
		PSOJ036+03    &$1.26_{-0.55}^{+0.77}$ & $<3.53$&$2.49_{-0.95}^{+1.27}$ & $8.20_{-3.57}^{+5.05}$ &$-1.77_{-0.10}^{+0.08}$ & $-0.05_{-0.10}^{+0.08}$  \\
		VIKJ0305$-$3150    & $<0.59$ & $<1.73$ & $<1.31$ &$<3.79$ & $<-1.72$ &$<-0.06$   \\
		SDSSJ0842+1218    & $0.75_{-0.38}^{+0.56}$& $<2.26$& $1.30_{-0.66}^{+0.94}$& $4.34_{-2.17}^{+3.26}$& $-1.81_{-0.12}^{+0.09}$& $-0.11_{-0.12}^{+0.09}$ \\
		PSOJ167$-$13    &$<0.32$& $<2.44$&$<1.39$&$<2.21$& $<-1.72$ & $<-0.09$   \\
		CFHQSJ1509$-$1749 &$1.67_{-0.62}^{+0.82}$ &$1.81_{-1.00}^{+1.47}$ &$2.28_{-0.91}^{+1.21}$ &$10.34_{-3.86}^{+5.10}$ &$-1.71_{-0.08}^{+0.07}$&$0.01_{-0.08}^{+0.07}$ \\
		CFHQSJ1641+3755  &$6.43_{-0.98}^{+1.07}$ & $2.85_{-0.93}^{+1.17}$&$10.65_{-1.49}^{+1.63}$&$33.39_{-5.07}^{+5.56}$ &  $-1.28_{-0.03}^{+0.03}$ &$0.35_{-0.03}^{+0.03}$\\
		PSOJ338+29     & $0.78_{-0.29}^{+0.39}$ & $<1.61$& $1.43_{-0.52}^{+0.66}$&  $5.92_{-2.22}^{+2.96}$& $-1.64_{-0.08}^{+0.07}$& $0.01_{-0.08}^{+0.07}$ \\
		SDSSJ2310+1855    & $1.22_{-0.59}^{+0.88}$& $<3.85$& $2.29_{-1.05}^{+1.49}$& $6.93_{-3.34}^{+5.02}$& $-1.87_{-0.11}^{+0.09}$& $-0.12_{-0.11}^{+0.09}$  \\
		\hline
		\multicolumn{7}{c}{{ Archival observations }} \\
		SDSSJ0100+2802&$7.28_{-0.47}^{+0.50}$ & $4.10_{-0.95}^{+1.03}$ & $14.09_{-0.91}^{+0.94}$ &$47.64_{-3.08}^{+3.27}$ & $-1.76_{-0.01}^{+0.01}$& $0.07_{-0.01}^{+0.01}$ \\
		ATLASJ0142$-$3327 & $2.26_{-0.68}^{+0.82}$ &	$<2.92$ &$3.75_{-1.14}^{+1.40}$ & $13.69_{-4.11}^{+4.98}$ &$-1.76_{-0.06}^{+0.05}$ & $-0.01_{-0.06}^{+0.05}$ \\
		CFHQSJ0210$-$0456  & $<4.83$ & $<5.162$ & $<6.62$ & $<29.31$ & $<-1.13$ & $<0.44$\\
		CFHQSJ0216$-$0455 & $<0.52$ &  $<0.92$ & $<0.70$ & $<2.70$ & $<-1.21$ & $<0.24$ \\
		SDSSJ0303$-$0019& $<15.9$ & $<108.5$ & $<46.00$ & $<97.70$ & $<-1.09$ & $<0.54$\\
		SDSSJ1030+0524 & $1.82_{-0.20}^{+0.21}$ & $2.05_{-0.30}^{+0.33}$ & $3.84_{-0.34}^{+0.36}$ & $10.77_{-1.18}^{+1.27}$ & $-1.68_{-0.02}^{+0.02}$ & $0.03_{-0.02}^{+0.02}$ \\
		SDSSJ1048+4637 &$0.77_{-0.37}^{+0.56}$ & $<2.66$  & $1.23_{-0.61}^{+0.97}$ & $4.25_{-2.05}^{+3.08}$ &   $-1.86_{-0.11}^{+0.09}$ & $-0.15_{-0.11}^{+0.09}$\\
		ULASJ1120+0641 & $0.73_{-0.15}^{+0.16}$ & $0.59_{-0.33}^{+0.39}$ & $1.41_{-0.30}^{+0.32}$ & $6.07_{-1.25}^{+1.33}$ & $-1.72_{-0.04}^{+0.03}$ & $-0.03_{-0.04}^{+0.03}$ \\
		SDSSJ1148+5251 & $2.17_{-0.40}^{+0.49}$ & $2.39_{-0.70}^{+0.87}$ & $4.51_{-0.70}^{+0.80}$ & $12.94_{-2.39}^{+2.69}$ & $-1.74_{-0.03}^{+0.03}$ & $-0.00_{-0.03}^{+0.03}$\\
		SDSSJ1306+0356 &	$3.25_{-0.31}^{+0.33}$ & $3.93_{-0.71}^{+0.81}$ & $6.93_{-0.60}^{+0.62}$ & $17.06_{-1.63}^{+1.73}$ &$-1.57_{-0.02}^{+0.02}$ & $0.12_{-0.02}^{+0.02}$\\
		ULASJ1342+0928    & $1.70_{-0.50}^{+0.62}$ & $1.82_{-0.80}^{+1.08}$ & $3.52_{-0.89}^{+1.06}$ & $14.96_{-4.40}^{+5.46}$ & $-1.57_{-0.06}^{+0.05}$ & $0.12_{-0.06}^{+0.05}$ \\
		SDSSJ1602+4228 &$7.09_{-1.39}^{+1.58}$ & $4.87_{-2.24}^{+3.16}$ & $13.05_{-2.50}^{+2.80}$ & $37.04_{-7.28}^{+8.25}$ & $-1.46_{-0.04}^{+0.03}$ & $0.24_{-0.04}^{+0.03}$ \\
		SDSSJ1623+3112 & $0.90_{-0.39}^{+0.55}$ & $2.92_{-1.41}^{+2.12}$ & $2.60_{-0.88}^{+1.15}$ & $5.08_{-2.22}^{+3.13}$ & $-1.73_{-0.10}^{+0.08}$ & $-0.05_{-0.10}^{+0.08}$\\
		SDSSJ1630+4012 & $1.92_{-0.50}^{+0.61}$ & $3.06_{-1.21}^{+1.66}$ & $4.35_{-0.97}^{+1.14}$ & $9.92_{-2.58}^{+3.12}$ &  $-1.57_{-0.05}^{+0.05}$ & $0.09_{-0.05}^{+0.05}$        \\	
		HSCJ2216$-$0016  & $<1.01$ & $<4.04$ & $<2.43$ & $<5.92$ & $<-1.25$ &$<0.27$ \\	
		\hline
	\end{tabular} \label{Tab_flux}\\
\end{table*}

\section{Results}

\subsection{$\alpha_{ox}$ vs. luminosity, redshift, and QSO properties}\label{aox}

We computed $L_{2500\angstrom}$ from the 1450 $\angstrom$ magnitude assuming a power-law spectrum with $\alpha=-0.3$, (e.g. \citealt{Banados16,Selsing16}). 
Tab.~\ref{Tab_flux} shows the $\alpha_{ox}$ values for the sources in our sample. The reported errors account only for the errors on the X-ray photometry, which dominate over the uncertainties on $L_{2500\angstrom}$. Errors on the UV luminosities are dominated by the assumed UV spectral slope rather than measurement errors. For instance, assuming $\alpha=-0.5$ \citep[e.g.][]{VandenBerk01} returns $\alpha_{ox}$ values steeper by \mbox{$\approx0.02$} than the reported ones, and thus still well within the errors on $\alpha_{ox}$ reported in Tab.~\ref{Tab_flux}. 

\begin{figure}
	\includegraphics[width=90mm,keepaspectratio]{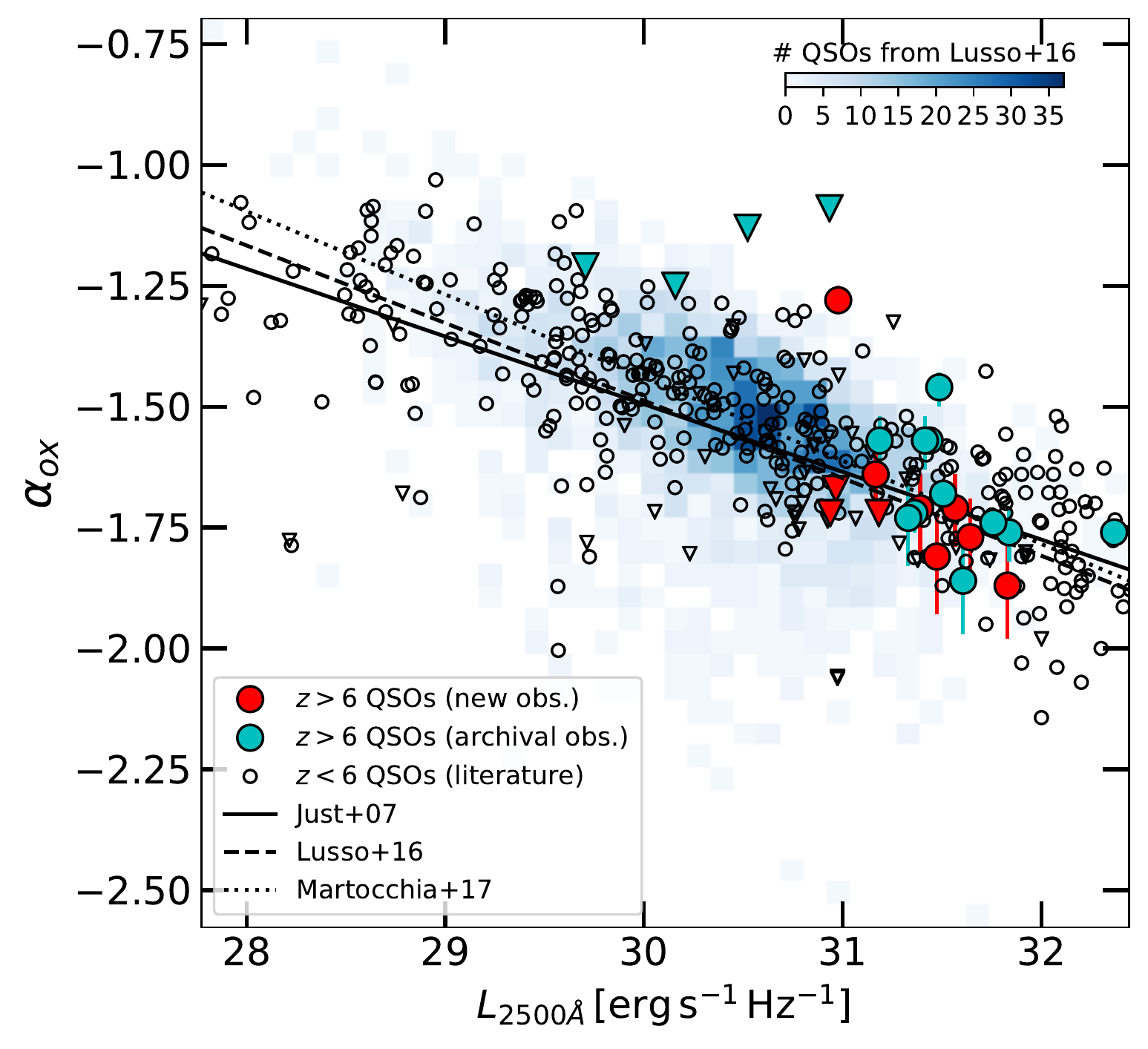} 
	\caption{$\alpha_{ox}$ vs. $L_{2500\angstrom}$ for $z\ge6$ QSOs, compared with a compilation of optically selected QSOs at lower redshifts \citep{Shemmer06b,Steffen06,Just07,Lusso16}. Downward pointing triangles represent upper limits. We also show the best-fitting relations of \citet{Just07}, \citet{Lusso16}, and \citet{Martocchia17}. For visual purposes, we do not plot X-ray undetected sources included in the \citet{Lusso16} sample. }
	\label{Fig_aox_L}
\end{figure}

We plotted in Fig.~\ref{Fig_aox_L} $\alpha_{ox}$ versus UV luminosity for our sample, and compared them with the best-fit relations of \cite{Just07}, \cite{Lusso16}, and \cite{Martocchia17}. All of these relations are very similar in the luminosity regime probed by our sources. We plot as small black symbols the sample of $z<6$ QSOs (from \citealt{Shemmer06}, \citealt{Steffen06}, and \citealt{Just07}) used to fit the \cite{Just07} relation. We also show the sample of  $>2000$ QSOs of \cite{Lusso16} as a color-coded map based on the number of sources per bin. For visual purposes only, we did not include upper limits (i.e. QSOs not detected in the X-rays) from \cite{Lusso16}, which would populate preferentially the steep $\alpha_{ox}$ regime.

In order to check if the $\alpha_{ox}$ values we found are in agreement with those expected from literature relations, we first note that the probability that a source is observed with an $\alpha_{ox}$ flatter or steeper than the expectation from a reference relation (we assumed the \citealt{Just07} one, based on optically selected QSOs as is our sample) due to random fluctuations only can be described by a binomial distribution, with probability of ``success" $p=0.5$ (i.e. we expect half of the sample to be above the relation), and number of trials $n=25$ (i.e. the sample size). Assuming the two extreme cases in which upper limits on $\alpha_{ox}$ are treated as a detection (i.e. $x=13$ sources above the relation) or represent sources \textit{intrinsically} below the relation (i.e. $x=9$), a binomial test returns probabilities of the observed or more extreme configurations given the expected configuration of $P=0.50$ and $P=0.11$, respectively. 
If we do not consider the four sources with weak upper limits on $\alpha_{ox}$, which do not provide useful information, we find $n=21$ and $x=9$, corresponding to $P=0.33$. According to these values, we do not find evidence supporting a significant variation of $\alpha_{ox}(L_{2500\angstrom})$ with redshift from this basic assessment.

\begin{figure*}
	\includegraphics[width=180mm,keepaspectratio]{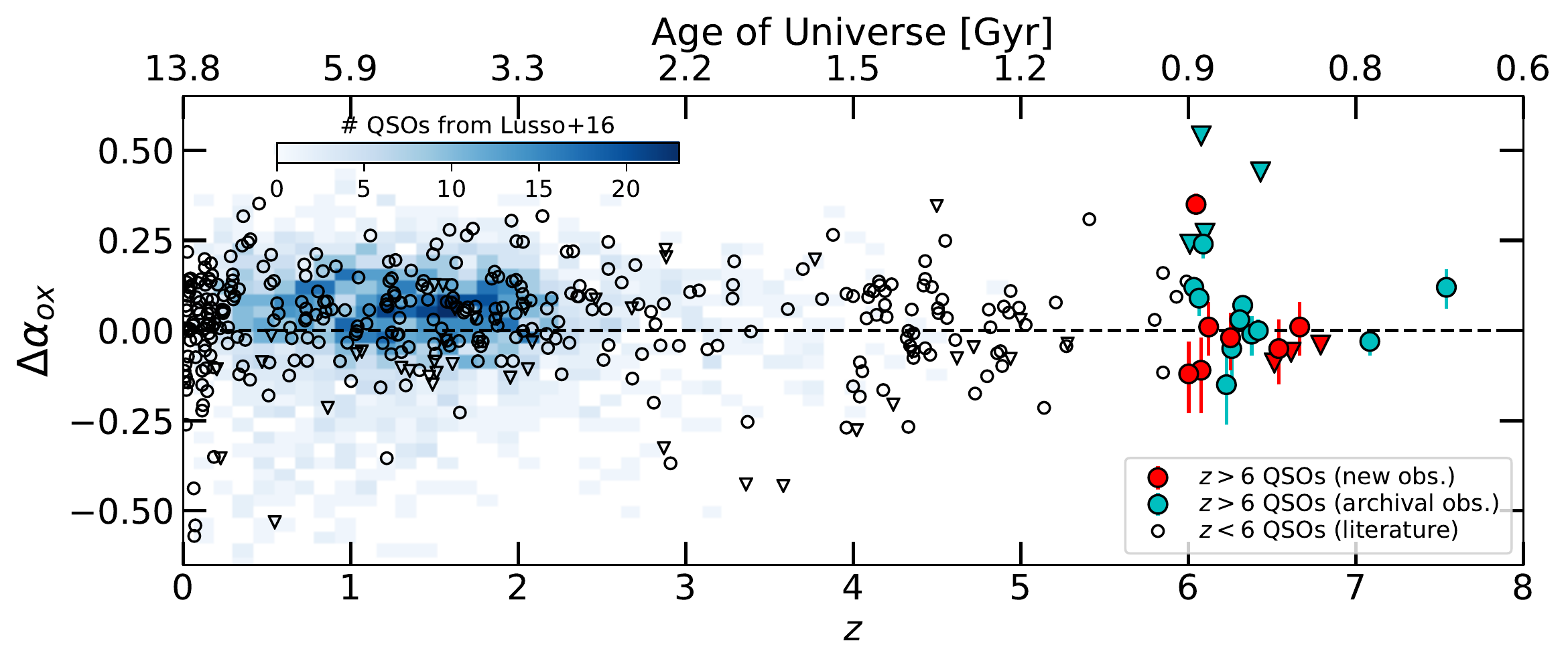} 
	\caption{$\Delta\alpha_{ox}$ vs. redshift for $z\ge6$ QSOs, compared with a compilation of QSOs at lower redshifts (see Fig.~\ref{Fig_aox_L}). Downward-pointing triangles represent upper limits. The horizontal dashed line corresponds to $\daox=0$.  }
	\label{Fig_daox_z}
\end{figure*}

This result can also be assessed by computing the difference between the observed $\alpha_{ox}$ and the value expected from the UV luminosity, according to the relation of \mbox{\cite{Just07}}, i.e. \mbox{$\Delta\alpha_{ox}=\alpha_{ox}(\mathrm{observed}) - \alpha_{ox}(L_{2500\angstrom})$} as a function of redshift (Fig.~\ref{Fig_daox_z}). 
If $\alpha_{ox}(L_{2500\angstrom})$ does not vary significantly with redshift, we expect the $\Delta\alpha_{ox}$ distribution of our sample to be consistent with the  distribution of the sample used by \cite{Just07} to fit their relation. We test this null hypothesis (i.e. that the two $\Delta\alpha_{ox}$ distributions are drawn from the same population) using the univariate methods \citep{Feigelson85} included in ASURV Rev. 1.2 \citep{Isobe90,Lavalley1992}, which allows accounting for censored data (i.e. sources undetected in X-rays). The null-hypothesis probabilities for the several tests we ran are reported in Tab.~\ref{Tab_tests}.
According to these tests, the $\Delta\alpha_{ox}$ distribution of our sample is consistent with those of lower-redshift samples collected from the literature. Finally, we computed the Kaplan-Meier estimator for the distribution function of the  $\Delta\alpha_{ox}$ parameters of the considered samples. Results are summarized in Tab.~\ref{Tab_KM}.\footnote{As reported in the ASURV manual, the Kaplan-Meier estimator requires the censoring to be random. Formally, this is not the case for our sample, as the censored variable, $\Delta\alpha_{ox}$, is directly related to the QSO luminosities, and less-luminous QSOs are more likely not to be detected. However, in addition to the luminosity of the QSOs, the censoring of $\Delta\alpha_{ox}$ is due to the flux limit of the observations (i.e. the exposure times) and the distances of the QSOs, which thus help to randomize the censoring distribution. } Following \citet[see their \S~3.5]{Steffen06}, we can estimate roughly the allowed fractional variation of the typical UV-to-X-ray flux ratio in QSOs as $\delta r/r= 2.606\, \mathrm{ln}(10)\alpha_{ox}\approx6\delta\alpha_{ox}=0.16$ at $1\sigma$, where $r=f_\nu(2500\,\angstrom)/f_\nu(2\,\mathrm{keV})$ and  $\delta\alpha_{ox}$ is the allowed variation of $\alpha_{ox}$, which we approximated with the uncertainty on the mean of $\Delta\alpha_{ox}$ computed with the Kaplan-Meier estimator. This estimate may be somewhat optimistic, as, for instance, we did not take into account the uncertainties on the \cite{Just07} $\alpha_{ox}-L_{2500\angstrom}$ relation.

\begin{figure}
	\includegraphics[width=90mm,keepaspectratio]{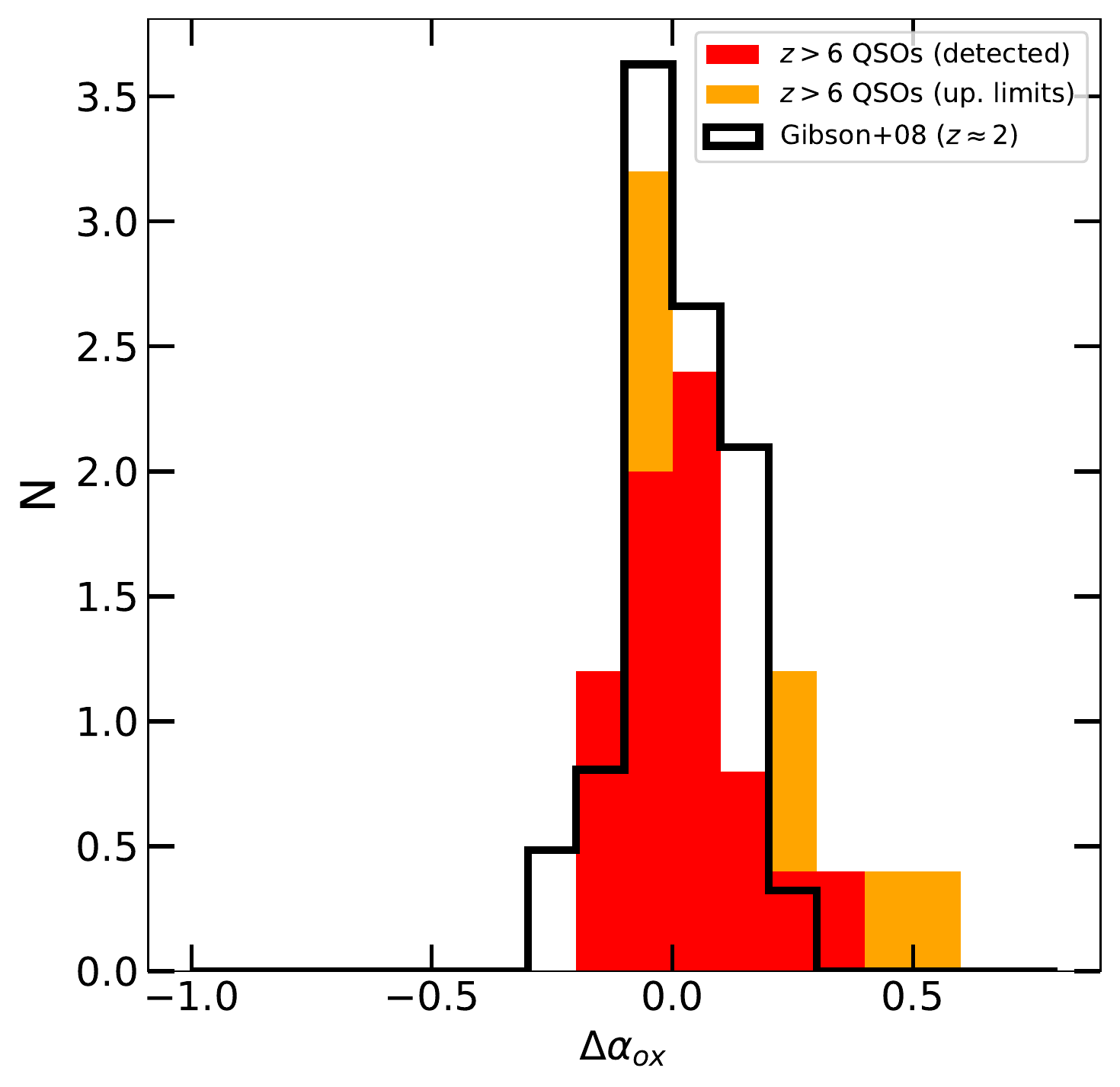} 
	\caption{Normalized histogram of $\Delta\alpha_{ox}$ for our sample of $z>6$ QSOs (detections and upper limits are represented with red and orange histograms, respectively), compared with the sample of $z\approx2$ QSOs presented as sample B of \citet{Gibson08}. }
	\label{Fig_daox_hist}
\end{figure}

In Fig.~\ref{Fig_daox_hist} we also compare the distribution of $\Delta\alpha_{ox}$ of our $z>6$ QSOs with the sample of $z\approx2$ QSOs presented in \citet[improved sample B; see footnote 3 of \citealt{Ni18}]{Gibson08}, which has been carefully selected to discard BALQSOs, and includes only X-ray detected QSOs. The two distributions are broadly consistent, again pointing toward a lack of a significant evolution of $\alpha_{ox}$ with redshift. We do not find a significant deviation of $\Delta\alpha_{ox}$ also limiting the tests to QSOs at the highest redshifts ($z>6.5$) in our sample, although we note that the size of such a subsample is too small (7 QSOs, 3 of which undetected)  to derive strong conclusions.

Based on the apparently non-evolving QSO $L_{UV}-L_X$ relation across cosmic time, \cite{Risaliti19} recently proposed the use of QSOs up to $z\approx5$ as standard candles to infer cosmological parameters, finding evidence for a deviation from the concordance $\Lambda$CDM model. In this respect, since type Ia supernovae are detected up to $z\approx1.4$ only, QSOs are particularly useful in the distant universe.

We do not find evidence supporting a significant correlation between $\Delta\alpha_{ox}$ and $M_\mathrm{BH}$, bolometric luminosity, or $\lambda_{Edd}$: Spearman's test returned $\rho=-0.10$ and $P=0.66$, $\rho=0.04$ and $P=0.84$, and $\rho=0.24$ and $P=0.29$, respectively. Note that $\Delta\alpha_{ox}$ factors out the dependence of $\alpha_{ox}$ with UV luminosity, which also enters into the computation of $M_\mathrm{BH}$ and bolometric luminosity, and it is thus a better parameter to use when checking for any potential correlation with such quantities.

QSO emission variability is potentially a significant source of uncertainty affecting the derived values of $\alpha_{ox}$ and $\Delta\alpha_{ox}$ \citep[e.g.][]{Gibson12, Vagnetti13}. For instance, \cite{Shemmer05} detected X-ray flux variability of a factor of $\approx4$ for SDSSJ02310--728 at $z=5.41$ over a rest-frame period of $\approx73$ days. 
\cite{Nanni18} found evidence for strong variability affecting the emission of SDSSJ1030+0524 at $z=6.308$ (see also \citealt{Shemmer05}): its X-ray flux increased by a factor of $\approx2.5$ from an \xmm\, observation in 2003 to the 2017 \chandra\, dataset analysed in this work, corresponding to a variation of $\Delta\alpha_{ox}$ of $\pm0.16$. As also discussed in \S~\ref{photometry}, we do not find other similar cases among the few other QSOs covered by multiple observations.

\begin{table}
	\caption{Probabilities that the $\Delta\alpha_{ox}$ distributions (including censored values) of our sample and samples taken from the literature (\citealt{Just07} and \citealt{Lusso16}) are drawn from the same parent population.}
	\begin{tabular}{ccccccccc} 
		\hline
		\multicolumn{1}{c}{{ Test }} &
		\multicolumn{1}{c}{{ $P$}} \\
		\hline
		\multicolumn{2}{c}{{ Just et al. (2007) sample ($z<6$) }} \\
		\hline
		Gehan's generalized  Wilcoxon Test & \\
		(permutation variance)  & 0.31\\
		(hypergeometric variance) & 0.30\\	
		Logrank Test& 0.67\\	
		Peto and Peto generalized Wilcoxon Test & 0.32 \\
		Peto and Prentice generalized Wilcoxon Test & 0.30 \\	
		\hline
		\multicolumn{2}{c}{{ Lusso \& Risaliti (2016) sample ($z<6$) }} \\
		\hline
		Gehan's generalized  Wilcoxon Test & \\
		(permutation variance)  & 0.49\\
		(hypergeometric variance) & 0.49\\	
		Logrank Test& 0.96\\	
		Peto and Peto generalized Wilcoxon Test & 0.50\\
		Peto and Prentice generalized Wilcoxon Test & 0.50 \\	
		\hline
		
	\end{tabular} \label{Tab_tests}\\
\end{table}

\begin{table}
	\caption{Results of the Kaplan-Meier estimator for the distribution function of $\Delta\alpha_{ox}$ for our $z>6$ sample and lower-redshift samples from \citet{Just07} and \citet{Lusso16}.}
	\begin{tabular}{ccccccccc} 
		\hline
		\multicolumn{1}{c}{{ Mean }} &
		\multicolumn{3}{c}{{ Percentiles}} \\
		\multicolumn{1}{c}{{$\Delta\alpha_{ox}$ }} &		
		\multicolumn{1}{c}{{ $25\%$ }} &
		\multicolumn{1}{c}{{ $50\%$ }} &		
		\multicolumn{1}{c}{{ $75\%$ }} \\
		\hline
		\multicolumn{4}{c}{{ $z>6$ sample (this work) }} \\	
		$0.005 \pm 0.026$ & $-0.113$ &     $-0.016$ &      0.056 \\
		\hline
		\multicolumn{4}{c}{{ Just et al. (2007) sample ($z<6$) }} \\	
		$0.005 \pm 0.009$ & $-0.066$ &     $0.033$ &      0.107 \\ 
		\hline
		\multicolumn{4}{c}{{ Lusso \& Risaliti (2016) sample ($z<6$) }} \\	
		$-0.036 \pm 0.011$ & $-0.084$ &     $0.021$ &      0.101 \\ 
		\hline
	\end{tabular} \label{Tab_KM}\\
\end{table}

\subsection{Bolometric corrections}\label{Kbol}
\begin{figure}
	\includegraphics[width=90mm,keepaspectratio]{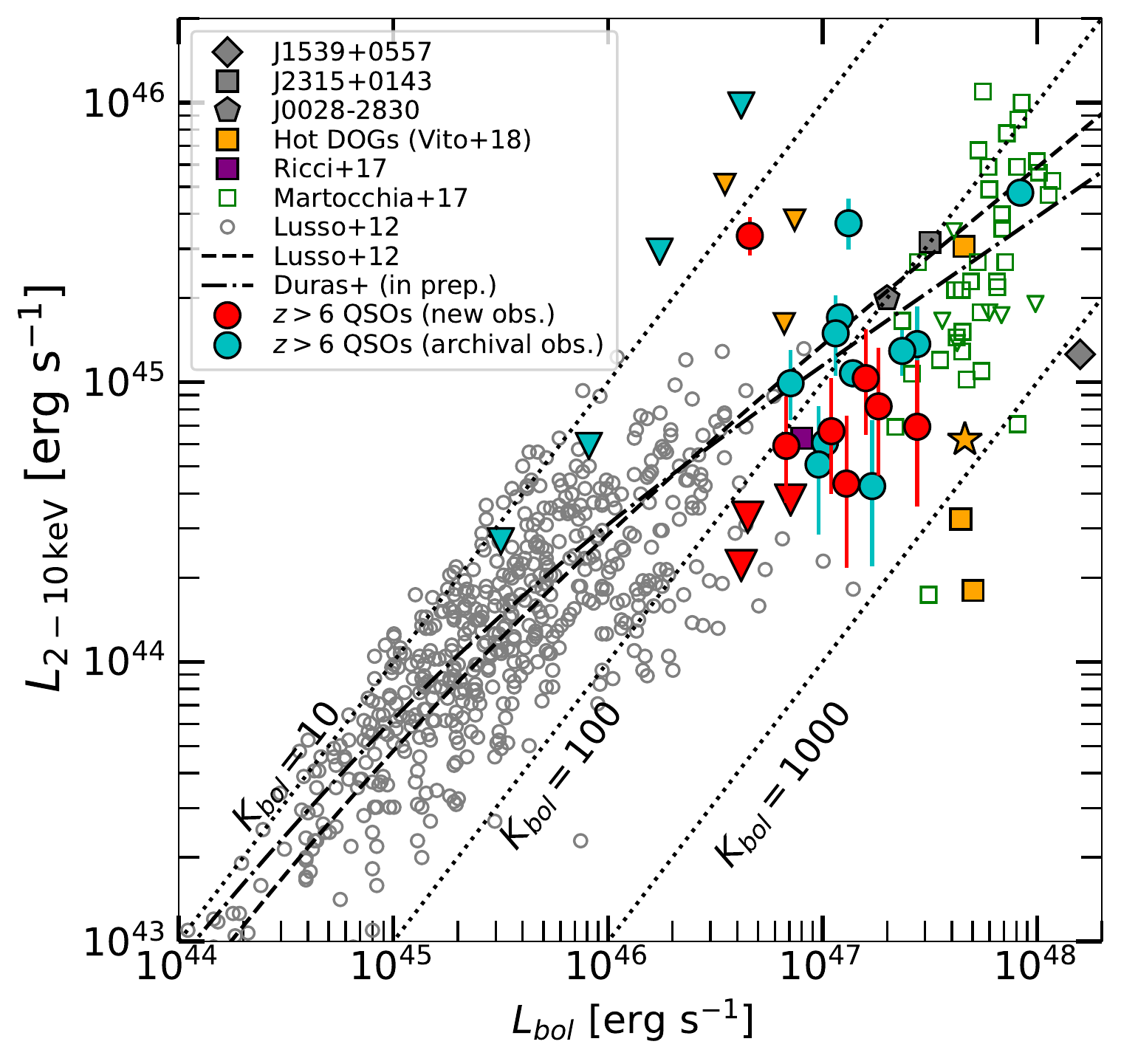} 
	\caption{X-ray versus bolometric luminosity of our sample of $z>6$ QOSs (red and cyan symbols), compared with the compilation of lower luminosity QSOs of \citet[empty grey circles]{Lusso12}. We also add the sample of luminous QSOs from \citet[green symbols]{Martocchia17}, the Hot DOG samples of \citet[orange symbols; the orange star represents the stacked result of several Hot DOGs undetected in the X-rays]{Vito18b} and \citet[purple square]{Ricci17}, and results for some individual hyperluminous QSOs \citep[filled grey symbols; see \citealt{Martocchia17} for their luminosities]{Feruglio14,Banerji15,Cano-Diaz12}. The dashed and dash-dotted black curves are the best-fitting relation of \citet{Lusso12} and Duras et al. (in prep.), respectively.
		Downward pointing triangles represent upper limits. Diagonal dotted lines mark the loci of constant bolometric correction. }
	\label{Fig_Lx_Lbol}
\end{figure}

Fig.~\ref{Fig_Lx_Lbol} presents the X-ray luminosities of $z>6$ QSOs plotted against their bolometric luminosities. We compare these with the sample of lower-luminosity Type 1 AGN selected in the XMM-COSMOS survey of \cite{Lusso10}, and with QSO samples with luminosities similar to or larger than those of our high-redshift sample \citep{Feruglio14, Banerji15, Cano-Diaz12, Martocchia17,Ricci17,Vito18b}. In particular, our sample populates a luminosity regime in this plane poorly sampled before.
The positions of our $z>6$ sources confirm the trend of increasing bolometric correction $K_{bol}=L_{bol}/L_X$ with bolometric luminosity, from $K_{bol}\approx10-100$ at log$L_{bol}\lesssim46.5$ to $K_{bol}\approx100-1000$ at log$\frac{L_{bol}}{\mathrm{erg\,s^{-1}}}\gtrsim46.5$, in agreement with previous works.
We note that the bolometric luminosities of our type 1 QSOs are derived from the UV luminosities as described in \S~\ref{other_properties}, with typical relative uncertainties of $\sim7\%$. Thus, the bolometric corrections found are byproducts of the relation shown in Fig.~\ref{Fig_aox_L}. 

\subsection{Spectral analysis }\label{spec_analysis}
\subsubsection{Individual sources}\label{spec_analysis_ind}
We performed a basic spectral analysis for individual sources in our sample, considering only those detected in at least one energy band, in order to compare the resulting parameters with those derived from hardness-ratio and aperture photometry analyses (Tab.~\ref{Tab_photometry} and Tab.~\ref{Tab_flux}). 
Spectra, response matrices, and ancillary files were extracted as described in \S~\ref{data_reduction}. We fitted the spectra with XSPEC v12.9.0n \mbox{\citep{Arnaud96}}.\footnote{\url{https://heasarc.gsfc.nasa.gov/xanadu/xspec/}} We used the $W$-statistic,\footnote{\url{https://heasarc.gsfc.nasa.gov/xanadu/xspec/manual/XSappendixStatistics.html}} which extends the \cite{Cash79} statistic in the case of background-subtracted data. In the case of a source observed by more than one instrument, we performed a joint spectral analysis using all the available spectra. Due to the generally limited photon counting statistics, we assumed a simple power-law model, and included Galactic absorption along the line of sight of each source \citep{Kalberla05}. The photon index and the power-law normalization are the only free parameters. Notwithstanding the simplicity of the model, the fit does not converge for SDSSJ0842+1218 (which has $\approx3$ net counts), which is thus not considered hereafter. For two other sources, PSOJ036+03, SDSS1048+5251, and SDSSJ2310+1855, the fit converges but returns only an upper limit on the power-law normalization, and thus on flux and luminosity.

Best-fit parameters are reported in Tab.~\ref{Tab_spec_analyis}.  
Although the uncertainties are typically large, the results derived from spectral and hardness-ratio analyses are consistent, suggesting that the procedures used in the previous sections are robust. The luminosity derived for PSOJ338+29 from spectral analysis is significantly larger than the value found from photometric analysis (Tab.~\ref{Tab_flux}), where we assumed $\Gamma=2.0$. This is due to the extremely steep best-fitting photon index derived from spectral analysis, likely due to the limited photon-counting statistics.

\begin{table}\small
	\caption{Best-fitting parameters derived from spectral analysis of individual sources (see \S~\ref{spec_analysis_ind}). Errors correspond to the 90\% confidence level for one parameter of interest \citep{Avni76}. }
	\begin{tabular}{ccccccccc} 
		\hline
		\multicolumn{1}{c}{{ ID }} &
		\multicolumn{1}{c}{{ $\Gamma$}} &
		\multicolumn{1}{c}{{ $F_{0.5-2\mathrm{keV}}$}} &
		\multicolumn{1}{c}{{ $L_{2-10\mathrm{keV}}$}}  \\
		\multicolumn{1}{c}{{}} &
		\multicolumn{1}{c}{{}} &
		\multicolumn{1}{c}{[$10^{-15}\,\funit$]} &
		\multicolumn{1}{c}{[$10^{44}\,\lunit$]}		\\	
		
		\hline
		\multicolumn{4}{c}{{ New observations }} \\
		CFHQSJ0050+3445  & $2.12_{-1.17}^{+2.01 }$ &     $1.45_{-0.89}^{+1.57}$ &  $8.20_{-5.03}^{+8.88}$ \\
		PSOJ036+03 & $2.10_{-1.50}^{+2.23 }$ &    $<3.05$ & $<20.53$\\
		CFHQSJ1509-1749    &$1.73_{-1.10}^{+1.29}$ &    $1.42_{-0.91}^{+1.72}$ & $7.69_{-4.93}^{+9.31}$\\
		CFHQSJ1641+3755$^*$   & $2.36_{-0.47}^{+0.50}$ &    $6.36_{-1.81}^{+2.26} $ & $39.20_{-11.21}^{+ 13.90}$\\
		PSOJ338+29 & $4.52_{-2.12}^{+2.57 }$ &    $1.41_{-0.83}^{+ 1.30}$ &  $57.31_{-33.64}^{+ 53.18}$\\
		SDSSJ2310+1855  & $3.18_{-3.67}^{+2.65 }$ &    $<3.34$ &  $<35.85$\\
		\multicolumn{4}{c}{{ Archival observations }} \\
		SDSSJ0100+2802$^*$ & $2.52_{-0.22}^{+0.23 }$ &   $7.71_{-1.02}^{+ 1.10}$ &  $67.55_{-8.93}^{+9.63}$\\
		ATLASJ0142-3327  & $2.03_{-1.10}^{+1.28 }$ &     $1.98_{-1.03}^{+1.43}$ &  $12.14_{-6.37}^{+8.52}$\\
		SDSSJ1030+0524$^*$ & $1.83_{-0.28}^{+0.29 }$ &    $1.76_{-0.38}^{+ 0.44}$ &  $9.55_{-2.04}^{+ 2.41}$\\
		SDSS1048+5251 & $1.84_{-1.56}^{+1.88 }$ &    $<1.52$ & $<7.79$\\
		ULASJ1120+0641$^*$  &$2.08_{-0.64}^{+0.74  }$ &  $0.68_{-0.28}^{+0.48} $ & $6.56_{-3.27}^{+3.59}$\\ 
		SDSSJ1148+5251$^*$  &$1.65_{-0.48}^{+0.50}$ &    $1.96_{-0.64}^{+0.83}$ &  $9.78_{-3.21}^{+4.11}$\\
		SDSSJ1306+0356$^*$ & $1.83_{-0.25}^{+0.26}$ &   $3.22_{-0.49}^{+5.44}$ &  $15.60_{-2.38}^{+2.64}$\\
		ULASJ1342+0928 &$1.97_{-0.92}^{+1.16 }$ &    $1.73_{-0.88}^{+1.33}$ & $14.95_{-7.60}^{+11.51}$\\
		SDSSJ1602+4228& $2.19_{-0.61}^{+0.74 }$ &   $ 6.89_{-2.10}^{+ 2.62}$ & $39.43_{-12.03}^{+14.99}$\\
		SDSSJ1623+3112  &$0.91_{-1.03}^{+2.40 }$ &  $0.89_{-0.59}^{+1.07}$ & $3.00_{-1.99}^{+ 3.62}$\\    
		SDSSJ1630+4012 &$1.90_{-0.69}^{+0.92 }$ &  $2.04_{-7.87}^{+ 1.05}$ & $10.03_{- 3.87}^{+ 5.15}$\\
		\hline
	\end{tabular} \label{Tab_spec_analyis}\\
	$^*$ These sources have $>30$ net counts in the $0.5-7$ keV band.
\end{table}

\subsubsection{Joint spectral analysis}\label{spec_analysis_joint}
We performed a joint spectral analysis to estimate the average photon index of sources detected in at least one energy band (18 sources). We removed the 6 QSOs with a total of more than 30 net counts in their spectra, for which results from individual spectral fitting are reported in \S~\ref{spec_analysis_ind}, as they would dominate the spectral-fit results. We used a single power-law model with photon index free to vary, but linked among the datasets, to fit jointly the remaining 12 sources ($\approx115$ net counts in the $0.5-7$ keV band) and added Galactic absorption appropriate to each source. 
We found a best-fitting, average photon index $\Gamma=2.20_{-0.34}^{+0.39}$ (errors at the 90\% c.l. corresponding to $\Delta W=2.7$; $\Gamma=2.20_{-0.20}^{+0.22}$ with errors at the 68\% c.l. corresponding to $\Delta W=1.0$).\footnote{The average derived through joint spectral analysis is by construction weighted by the number of counts of each spectrum, and thus depends on a complex combination of source fluxes and exposure times.} Repeating the joint spectral analysis for the 6 QSOs with $>30$ net counts ($\approx746$ net counts in total), we found an average $\Gamma=2.13_{-0.13}^{+0.13}$ ($\pm0.08$ at the the 68\% c.l.). Considering only QSOs at $z>6.5$, ULAS1120+0641 is detected with $>30$ counts, and its best-fitting photon index is $\Gamma\approx2$ (see Tab.~\ref{Tab_spec_analyis}). Joint spectral analysis of the other three $z>6.5$ QSOs detected in the X-rays ($\approx23$ net counts in total) returns $\Gamma=2.66_{-0.78}^{+0.92}$ ($\Gamma=2.66_{-0.50}^{+0.54}$ with errors at the 68\% c.l.).

\begin{figure}
	\includegraphics[width=90mm,keepaspectratio]{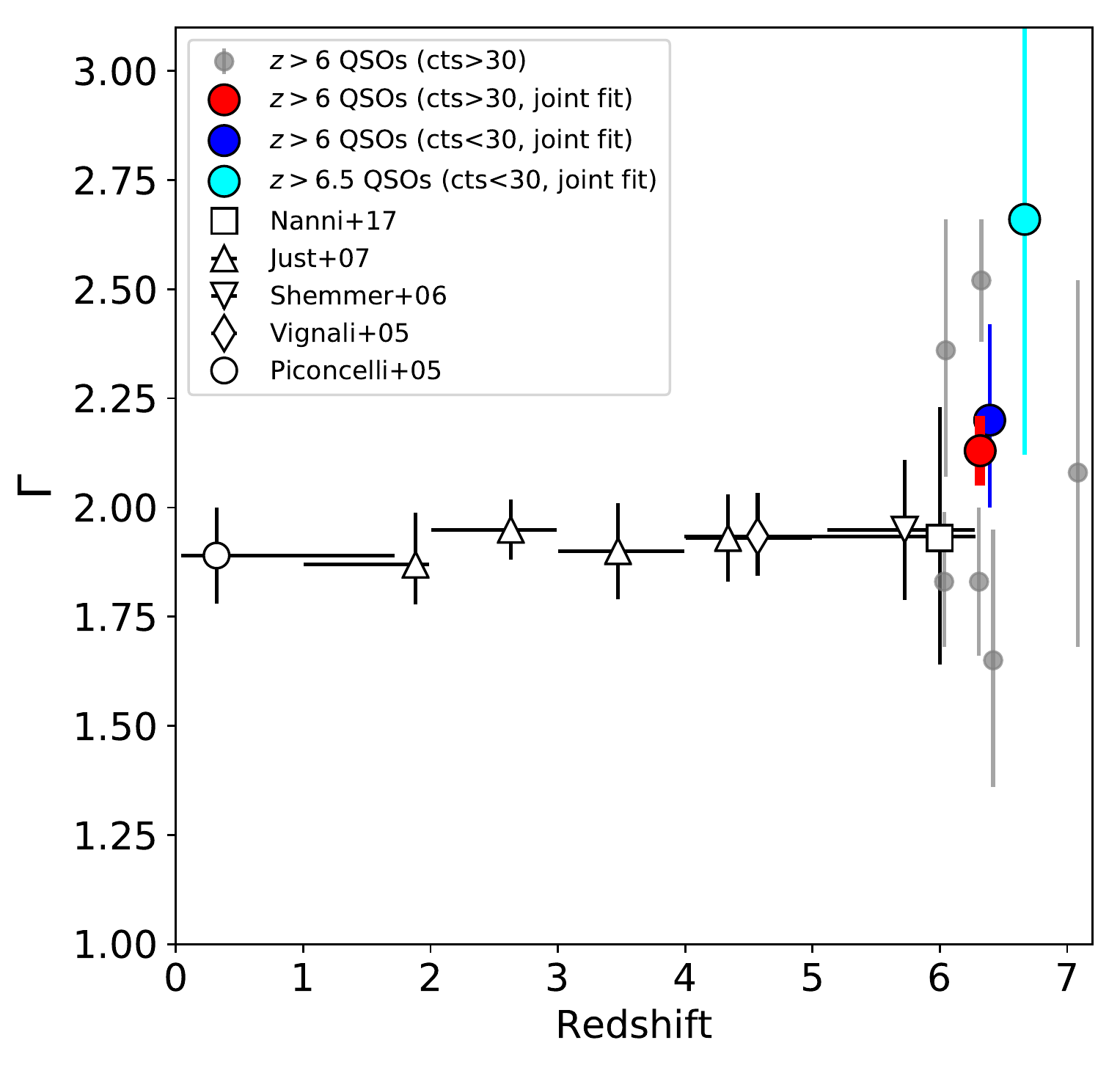} 
	\caption{Photon index as a function of redshift. We report the individual best-fitting values for sources with $>30$ total net counts (grey symbols), the results derived from joint spectral analysis of QSOs with $>30$, $<30$ net counts, and of $z>6.5$ QSOs (red, blue, and cyan circles, respectively, plotted at the median redshift of each subsample), and the average photon indices derived by \citet{Piconcelli05}, \citet{Vignali05}, \citet{Shemmer06}, \citet{Just07}, and \citet{Nanni17} for optically selected luminous QSOs at different redshifts. Errors are at the 68\% confidence level.
	}
	\label{Fig_Gamma_z_L}
\end{figure}

All of these values are slightly steeper than, although still consistent with, that found by \cite{Nanni17} for $z>5.7$ QSOs, thus including a subsample of our sources (i.e. $\Gamma=1.93_{-0.29}^{+0.30}$), and with the findings of \mbox{\cite{Piconcelli05}},  \mbox{\cite{Vignali05}}, \cite{Shemmer06b}, and \cite{Just07} at lower redshifts (Fig.~\ref{Fig_Gamma_z_L}). Thus, we conclude there is no strong evidence supporting a significant systematic variation of $\Gamma$ in our sample, although there are hints of a steepening of the typical QSO photon index at $z>6$.

The observed-frame $0.5-7$ keV band corresponds to rest-frame energies at $z>6$ where  a possible Compton-reflection component would peak in the X-ray spectra of QSOs. We did not account for this component in the spectral fitting, due to the small number of total counts preventing the use of relatively complex models. However, we note that the reflection component in the X-ray spectra of luminous Type-1 QSOs has been found to be generally weak both in the local universe \citep[e.g.][]{Comastri92, Piconcelli05} and at high redshift \citep[$z>4$, e.g.][]{Shemmer05}. Moreover, a strong reflection component would tend to flatten systematically the observed effective photon index, in contrast with our results.

Performing joint spectral analysis on subsamples of QSOs divided on the basis of their Eddington ratios, we do not find any significant trend of $\Gamma$ with $\lambda_{Edd}$. However, this may be due to the small sample size, and the large uncertainties affecting the single-epoch black hole masses and the best-fitting photon indices.

\begin{figure}
	\includegraphics[width=70mm,keepaspectratio, angle=270,trim={0cm 1cm 0cm 3cm}]{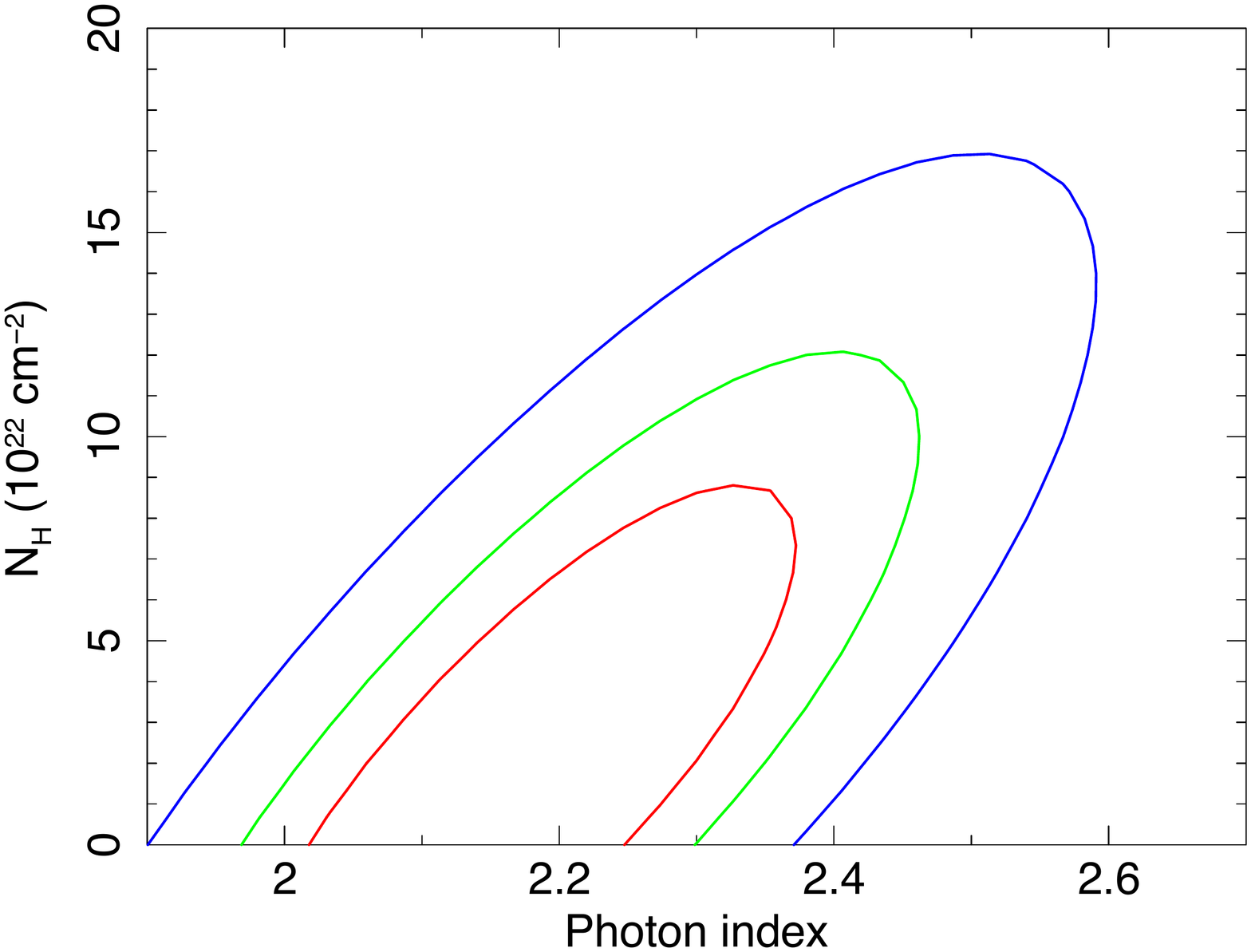} 
	\caption{Confidence contours at 68\%, 90\%, and 99\% confidence levels (red, green, and blue curves, respectively) of the best-fitting column density and photon index derived from a joint spectral analysis of the QSOs with $>30$ counts (see \S~\ref{spec_analysis_joint}).
	}
	\label{Fig_Gamma_NH}
\end{figure}

In order to place a basic upper limit on the average column density, we added an XSPEC \textit{zwabs} component and repeated the joint fit of QSOs with $>30$ net counts. We left both the photon index and the column density free to vary, but linked them among the spectra, and fixed the redshift to the appropriate value for each QSO. The best-fitting parameters are $\Gamma=2.17_{-0.14}^{+0.22}$ and $N_H<9\times10^{22}\,\mathrm{cm^{-2}}$ at the 90\% confidence level (see Fig.~\ref{Fig_Gamma_NH} for the confidence contours). The upper limit on $N_H$ is dominated by the high-redshift nature of the sources, which causes the photoelectric cutoff to shift below \chandra\, observed energy bands even for possible moderately high values of column density.

\subsection{Comments on  individual QSOs}\label{ind_sources}
\subsubsection{PSOJ167--13}
This QSO ($z=6.515$) falls slightly below our detection threshold in the hard band ($P_B=0.989$). We then checked whether we could detect it by restricting the detection energy range to  the $2-5$ keV band. This choice is motivated by the drop of the \chandra\, effective area and the relatively high background level at higher energies. Moreover, observed energies $E>5$ keV correspond to $E>37.5$ keV in the QSO rest frame, where the number of emitted X-ray photons is limited due to the QSO power-law spectrum.

\begin{figure}
	\includegraphics[width=90mm,keepaspectratio]{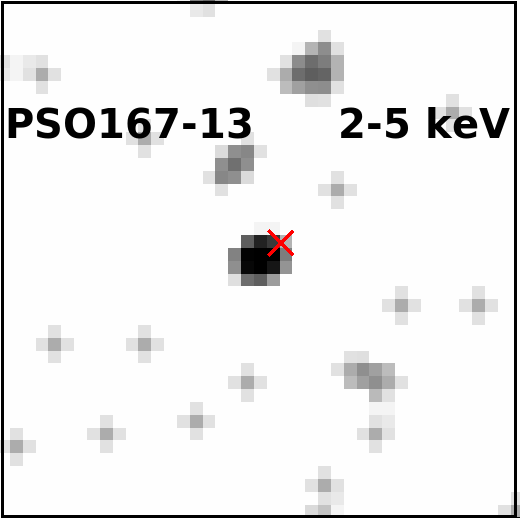} 

	\caption{Smoothed $2-5$ keV image ($40\times40$ pixels; $\approx20''\times20''$) of PSO167--13. The red cross marks the optical position of the QSO.
	}
	\label{Fig_PSO167m13}
\end{figure}

An X-ray source is significantly detected (\mbox{$P_B=4\times10^{-4}$}) with $2.9_{-1.4}^{+2.1}$ net counts in the \mbox{$2-5$ keV} band in an $R=1$ arcsec circular region (Fig.~\ref{Fig_PSO167m13}). The centroid of the X-ray emission shows an offset of $\approx1$ arcsec with respect to the optical position of the QSO, but with a positional uncertainty of $1.2$ arcsec at the 90\% confidence level. Considering the lack of counts detected in the soft band, following the procedure used in \S~\ref{photometry}, we derived $HR>0.47$ and $HR>0.08$ at the 68\% and 90\% confidence levels, corresponding to $\Gamma_{eff}<0.55$ and $\Gamma_{eff}<1.54$, respectively. This very hard spectrum at $z=6.515$, assuming an intrinsic $\Gamma=2$ spectrum, corresponds to lower limits on the obscuring column density of $N_H>2\times10^{24}\,\mathrm{cm^{-2}}$ and $N_H>6\times10^{23}\,\mathrm{cm^{-2}}$ at the $68\%$ and $90\%$ confidence levels, respectively. Therefore this object is the first heavily obscured QSO candidate at $z>6$, with the intriguing property of being an optically classified Type 1 QSO.

An ALMA sub-mm observation \citep{Willott17} revealed the presence of a close galaxy companion from the rest-frame UV and [C II] position of the QSO ($0.9$ arcsec; i.e. $\approx5$ kpc in projection at the redshift of the QSO), and by $\Delta v\approx-270\,\mathrm{km\,s^{-1}}$ (i.e. $\Delta z \approx 0.007$) in velocity space. The offset between the X-ray centroid and the [C II] position of this galaxy is only $\approx0.15$ arcsec. A thorough investigation and discussion of this system has been presented separately \citep{Vito19}. 

\subsubsection{VIK0305--3150}
\begin{figure}
	\includegraphics[width=90mm,keepaspectratio]{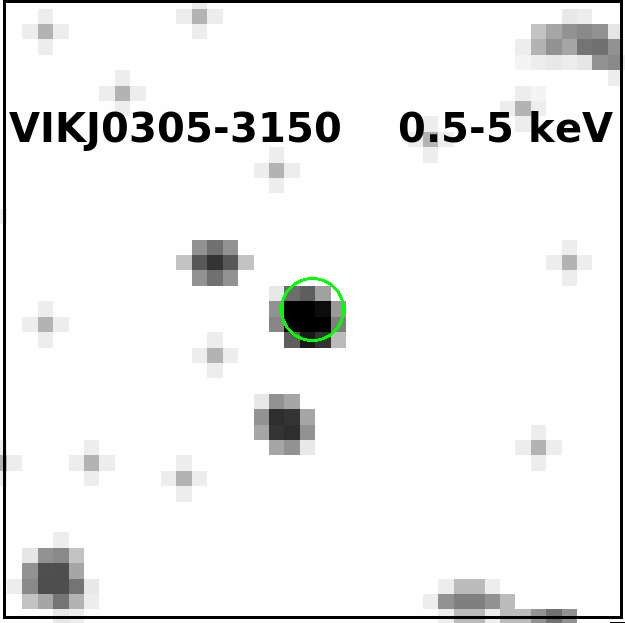} 

	\caption{Smoothed $0.5-5$ keV image ($40\times40$ pixels; $\approx20''\times20''$) of VIKJ0305--3150. The green circle has a radius $R=1$ arcsec.
	}
	\label{Fig_VIKJ0305}
\end{figure}

Similarly to PSO167--13, VIK0305--3150 ($z=6.047$) is slightly below our detection threshold both in the hard and full bands. We thus repeated the analysis restricting the energy bands to $2-5$~keV and $0.5-5$~keV. We nominally detected this QSO in a $R=1$ arcsec circular region in the $2-5$~keV band with $P_B=6.6\times10^{-3}$, but with a very limited number of net counts ($1.9^{+1.8}_{-1.1}$). The detection in the $0.5-5$~keV band is more solid ($P_B=1.1\times10^{-3}$) with $2.8^{+2.1}_{-1.4}$ net counts (Fig.~\ref{Fig_VIKJ0305}). Repeating the same hardness-ratio analysis as done above for PSO167--13, we found $HR>0.00$ and $HR>-0.39$, corresponding to $\Gamma_{eff}<0.96$ and $\Gamma_{eff}<1.93$ at the 68\% and 90\% confidence levels, respectively. Assuming an intrinsic $\Gamma=1.9$, the nominal obscuring column density is $N_H>1\times10^{24}\,\mathrm{cm^{-2}}$ at the 68\% confidence level, but it is not constrained at the 90\% confidence level.

\subsubsection{CFHQSJ1641+3755}

This radio-quiet ($R<10.5$) QSO at $z=6.047$ has one of the lowest bolometric luminosities (log$\frac{L_{bol}}{L_\odot}=13.1$) and smallest black hole masses (log$\frac{M_{BH}}{M_\odot}=8.4$) among the $z>6$ QSO sample, resulting in a high Eddington ratio  ($\lambda_{Edd}=1.5$). While the bolometric correction is usually found to anti-correlate with the Eddington ratio both observationally \citep[e.g.][]{Lusso12} and theoretically \citep[e.g.][but see also \citealt{CastelloMor17}]{Meier12, Jiang17}, CFHQSJ1641+3755 is the second most X-ray luminous $z>6$ QSO, resulting in a bolometric correction $K_{bol}\approx13$. This is also reflected in a quite flat $\alpha_{ox}=-1.28$. Considering an \mbox{$rms=0.2$} for the $\alpha_{ox}-L_{UV}$ relation in the $L_{UV}$ luminosity range where CFHQSJ1641+3755 lies \citep{Steffen06}, this QSO is a $\approx1.8\sigma$ outlier. 

Its $\approx50$ net counts allowed us to constrain with a reasonable accuracy its photon index, which, as expected considering its high Eddington ratio \citep[e.g.][]{Brightman13, Fanali13}, is quite steep ($\Gamma=2.36_{-0.47}^{+0.50}$). Consistent results for the estimated photon index and X-ray luminosity are found from the photometric and spectral analyses (see Tab.~\ref{Tab_photometry}, Tab.~\ref{Tab_flux}, and Tab.~\ref{Tab_spec_analyis}).

\subsubsection{WLQs and BALQSOs}
As discussed at the end of \S~\ref{other_properties}, among our sample are BALQSO and WLQ candidates, which are often associated with weak X-ray emission. 
The UV spectral quality of several of these objects prevents us from securely including them in one of these classes, but we note that the two QSOs with the most negative $\Delta\alpha_{ox}$ values (Tab.~\ref{Tab_flux}) are a WLQ candidate (J2310+1855) and a known BALQSO (J1048+4637). The remaining candidates have $\Delta\alpha_{ox}$ values consistent with the rest of the sample. Detected WLQ and BALQSO candidates do not show particularly flat photon indices (see Tab.~\ref{Tab_spec_analyis}), which might suggest the presence of a significant level of X-ray absorption.

\section{Conclusions}
We have presented new \chandra\, observations of 10 $z>6$ QSOs, selected to be radio quiet and to have virial black-hole mass estimates from Mg II line measurements. With this sample, we more than triple the number of QSOs at $z>6.5$ with existing sensitive \mbox{X-ray} coverage. In particular, five of the targets have UV magnitudes $-26.2<M_{1450\angstrom}<-25.6$, and are thus the least luminous $z>6$ QSOs targeted with sensitive X-ray observations. We detected 7/10 of our new targets in at least one standard energy band, and 2 additional QSOs discarding $E>5$ keV.
Adding archival observations at $z>6$, we could study the X-ray properties of a statistically significant sample of 25 QSO in the first Gyr of the universe. Our main results are the following:

\begin{itemize}

	\item From photometric analysis, we constrained or derived upper limits on the X-ray luminosity of $z>6$ QSOs, and their basic spectral shape, modeled with a simple power law. Consistent results are found from spectral analysis of individual bright sources, although the derived individual best-fitting photon indices have large uncertainties. See \S~\ref{photometry} and \ref{spec_analysis}.
	
	\item We do not find evidence for a significant evolution of the relation between QSO UV and X-ray luminosity, as traced by the $\alpha_{ox}$ parameter. The luminosities of $z>6$ QSOs are consistent with relations found at lower redshift for optically selected QSOs \citep[e.g.][]{Just07, Lusso16}, implying that the coronal emission becomes less important compared with disk emission at high luminosity also at $z>6$. 
	See \S~\ref{aox}.
	
	\item We do not find significant correlations between $\alpha_{ox}$ and black-hole mass or Eddington ratio, once the dependence of all of these quantities with the QSO UV luminosity is taken into account. See \S~\ref{aox}.
	
	\item We confirm the trend of increasing bolometric correction with increasing luminosity from $K_{bol}\approx10-100$ at log$\frac{L_{bol}}{L_{L_\odot}}\lesssim46.5$ to $K_{bol}\approx100-1000$ at log$\frac{L_{bol}}{L_{L_\odot}}\gtrsim46.5$, for the first time at $z>6$.
	In particular, our sample populates the luminosity region between moderate luminosity QSOs and ultra-luminous QSOs, currently poorly sampled. See \S~\ref{Kbol}.

	\item We perform a basic spectral analysis of sources with $>30$ net counts, and derived typical photon indices \mbox{$\Gamma\approx1.6-2.5$}. Joint spectral analysis of fainter sources returned an average value ($\Gamma=2.13_{-0.13}^{+0.13}$ and $\Gamma=2.20_{-0.34}^{+0.39}$, for sources with $>30$ and $<30$ net counts, respectively) slightly steeper, but still consistent with, typical photon indices of lower redshift QSOs. This result again supports a scenario in which the accretion-disk/hot-corona structure does not evolve strongly from low redshift to $z>6$. See \S~\ref{spec_analysis}.
	
	\item  Two of the three undetected targets could be detected by restricting the energy range to avoid background-dominated regions ($E>5$ keV). In particular, one of these, PSO167--13, presents a very hard spectrum, consistent with a large obscuring column density, and it is thus the first heavily obscured QSO candidate at $z>6$. See \S~\ref{ind_sources}.
	
\end{itemize}
Only $\approx25$ of the $z>6$ QSOs have been currently observed in the X-rays, while the number of known high-redshift QSOs is continuously growing. Moreover, over the coming $\approx10-20$ years, wide-field surveys (e.g. \textit{Euclid}, \textit{eROSITA}, \textit{LSST}, \textit{SUMIRE-HSC}, and \textit{WFIRST}) are expected to push the QSO redshift frontier far into the reionization era, detecting hundreds of accreting SMBHs at $z\approx 7-10$ \citep[e.g.][]{Brandt17}. Studying QSO properties in the first few $10^8$ years of the Universe will be extremely important to understand some of the major open issues in modern astrophysics, such as the formation and early growth of SMBHs, their interplay with proto-galaxies, the formation of the first structures, and the mechanisms responsible for the reionization of the Universe. Observing larger samples of high-redshift QSOs with \textit{Chandra} and \textit{XMM-Newton} will provide key \mbox{X-ray} information on their small-scale accretion physics, even in the presence of heavy obscuration, and will pave the way for future X-ray observatories, such as \textit{Athena}, \textit{Lynx}, and \textit{AXIS}. It is especially important to assess if the
hints we find for steepening X-ray power-law spectra, and high associated
Eddington ratios, become stronger at still higher redshifts. Targeting of $z>8$ QSOs in the next decades will take advantage of the tightest constraints we have placed on the  X-ray properties of the $z\approx6-7$ QSO population. In particular, realistic exposure-time estimates can be computed on the basis of the lack of a strong evolution of the $L_X-L_{UV}$ relation up to the highest redshifts which can be probed currently.

\section*{Acknowledgments} \vspace{0.2cm}
We thank the anonymous referee for useful feedback. We thank L. Jiang for providing the near-IR spectrum of SDSSJ2310+1855, A. Moretti for his help in reducing Swift data, C. Willott for useful discussions, E. Picconcelli and S. Martocchia for providing bolometric luminosities of the WISSH QSOs, and F. Duras for providing their functional form of the bolometric correction curve.
FV acknowledges financial support from CONICYT and CASSACA through the Fourth call for tenders of the CAS-CONICYT Fund, CONICYT grants Basal-CATA AFB-170002 (FV, FEB), 
the Ministry of Economy, Development, and Tourism's Millennium Science
Initiative through grant IC120009, awarded to The Millennium Institute
of Astrophysics, MAS (FEB). WNB acknowledges support from CXC grant G08--19076X and NASA ADP grant 80NSSC18K0878. BL acknowledges financial support from the National 
Key R\&D Program of China grant 2016YFA0400702 and 
National Natural Science Foundation of China grant 
11673010. We acknowledge financial contribution from the agreement ASI-INAF n.2017-14-H.O.

\bibliographystyle{aa}
\bibliography{../../../../biblio.bib} 

%
\end{document}